\documentclass[usenatbib]{mn2e}
\usepackage{color}
\usepackage{graphicx}
\usepackage{amssymb}

\title[Stellar Populations of Lyman Alpha Emitters]
 {
Stellar Populations of Lyman Alpha Emitters at $z=3$ -- $4$\\
Based on Deep Large Area Surveys\\
in the Subaru-SXDS/UKIDSS-UDS Field
}
\author[Y. Ono et al.]
{Yoshiaki~Ono,$^1$
Masami~Ouchi,$^{2,3}$
Kazuhiro~Shimasaku,$^{1,4}$
Masayuki~Akiyama,$^5$
\newauthor
James~Dunlop,$^{6,7}$
Duncan~Farrah,$^{8,9}$
Janice C. Lee,$^{2,10}$
Ross~McLure,$^7$
\newauthor
Sadanori~Okamura,$^{1,4}$
and
Makiko~Yoshida$^1$
\\
$^1$
Department of Astronomy, Graduate School of Science,
The University of Tokyo, Tokyo 113-0033, Japan \\
$^2$
Observatories of the Carnegie Institution of Washington,
813 Santa Barbara Street, Pasadena, CA 91101, USA \\
$^3$
Carnegie Fellow \\
$^4$
Research Center for the Early Universe, Graduate School of Science,
The University of Tokyo, Tokyo 113-0033, Japan \\
$^5$
Astronomical Institute, Graduate School of Science,
Tohoku University, Aramaki, Aoba, Sendai 980-8578, Japan \\
$^6$
Department of Physics and Astronomy, University of British Columbia,
6224 Agricultural Road, Vancouver V6T 1Z1, Canada \\
$^7$
SUPA Institute for Astronomy, University of Edinburgh, Royal
Observatory, Edinburgh EH9 3HJ, UK \\
$^8$
Department of Astronomy, Cornell University, Ithaca, NY 14853 \\
$^9$
Astronomy Centre, University of Sussex, Falmer, Brighton, UK \\
$^{10}$
Hubble Fellow
}

\date{Accepted 2009 November 11. Received 2009 October 28; in original form 2009 May 20}

\pagerange{\pageref{firstpage}--\pageref{lastpage}} \pubyear{2009}

\begin{document}

\label{firstpage}

\maketitle

\begin{abstract}
We investigate the stellar populations of Lyman $\alpha$ emitters (LAEs)
at $z=3.1$ and $3.7$ in $0.65$ deg$^2$
of the Subaru/\textit{XMM-Newton} Deep Field,
based on rest-frame UV-to-optical photometry
obtained from the Subaru/\textit{XMM-Newton} Deep Survey,
the UKIDSS/Ultra Deep Survey,
and the Spitzer legacy survey of the UKIDSS/UDS.
Among a total of $302$ LAEs ($224$ for $z=3.1$ and $78$ for $z=3.7$),
only $11$ are detected in the $K$ band,
i.e., brighter than $K(3\sigma)=24.1$ mag.
Eight of the $11$ $K$-detected LAEs are spectroscopically confirmed.
In our stellar population analysis,
we treat $K$-detected objects individually,
while $K$-undetected objects are stacked at each redshift.
We find that the $K$-undetected objects,
which should closely represent the LAE population as a whole,
have low stellar masses of $\sim 10^8$ -- $10^{8.5} M_\odot$,
modest SFRs of $1$ -- $100$ $M_\odot$ yr$^{-1}$,
and modest dust extinction of $E(B-V)_\star < 0.2$.
The $K$-detected objects are massive,
$M_{\rm star} \sim 10^9$ -- $10^{10.5} M_\odot$,
and have significant dust extinction with a median of
$E(B-V)_\star \simeq 0.3$.
Four $K$-detected objects with the reddest spectral energy distributions,
two of which are spectroscopically confirmed,
are heavily obscured with $E(B-V)_\star \sim 0.65$,
and their continua resemble those of some local ULIRGs.
Interestingly, they have large Lyman $\alpha$ equivalent widths
$\simeq 70$ -- $250${\AA}.
If these four are excluded, our sample has a weak anti-correlation
between Ly$\alpha$ equivalent width and $M_{\rm star}$.
We compare the stellar masses and
the specific star formation rates (sSFR) of LAEs
with those of Lyman-break galaxies (LBGs), distant red galaxies, submillimetre galaxies,
and $I$- or $K$-selected galaxies with photometric redshifts
of $z_{\rm phot} \sim 3$.
We find that the LAE population is the least massive
among all the galaxy populations in question, but with
relatively high sSFRs, while NIR-detected LAEs have
$M_{\rm star}$ and sSFR similar to LBGs.
Our reddest four LAEs have very high sSFRs in spite of
large $M_{\rm star}$, thus occupying a unique region
in the $M_{\rm star}$ versus sSFR space.
\end{abstract}



\section{INTRODUCTION} \label{sec:intro}

Lyman $\alpha$ emitters (LAEs) are a galaxy population
which are common in the high redshift Universe.
In the last decade, many observations have succeeded in detecting
LAEs from $z \sim 2$ up to $z \sim 7$, primarily based on
narrow-band imaging to isolate Lyman $\alpha$ emission
\citep[e.g.][]{hu1998,rhoads2000,iye2006}.
Over a thousand LAEs are now photometrically or spectroscopically
identified
\citep[e.g.][]{hu2002,ouchi2003,malhotra2004,
taniguchi2005,shimasaku2006,kashikawa2006,
dawson2007,murayama2007,gronwall2007,ouchi2008}.

Most LAEs have relatively faint, blue UV continua and
large Ly$\alpha$ equivalent widths (EWs).
These properties collectively suggest that they are young star-forming
galaxies with low metallicities \citep[e.g.][]{malhotra2002}.
In this sense, they may serve as building blocks of larger galaxies
in hierarchical galaxy formation.

Stellar population analysis is helpful
to further constrain the nature of LAEs.
Recent multiwavelength observations covering near-infrared
wavelengths have enabled analysis of the stellar populations
of LAEs, and there is growing evidence that not all LAEs
are primordial as described above.
Table \ref{tab:summary_of_earliers}
summarises studies of the stellar populations of LAEs,
including the work presented here.
\cite{gawiser2007} have performed a stacking analysis of $52$ LAEs at $z=3.1$,
and concluded that LAEs have
low stellar masses ($\simeq 10^9 M_\odot$),
young-age components ($\simeq 20$ Myr),
and small dust extinctions ($A_V = 0$)
\citep[see also][]{gawiser2006, nilsson2007}.
\cite{pirzkal2007} have studied nine LAEs at $z \sim 5$ found by
HST/ACS slitless spectroscopy in the Hubble Ultra Deep Field,
to show that these faint LAEs are
all very young ($\simeq 1$ -- $20$ Myr)
with low masses ($\simeq 10^6$ -- $10^8 M_\odot$)
and small dust extinctions ($A_V = 0$ -- $0.6$).
On the other hand, \cite{lai2007} have concentrated on three
spectroscopically confirmed LAEs at $z = 5.7$ with IRAC detection,
and derived relatively large stellar masses,
$\simeq 10^9$ -- $10^{10} M_\odot$,
mild dust extinctions ($E(B-V) \simeq 0.15$ -- $0.2$),
and a wide range of age, $\simeq 5$ -- $100$ Myr.
\cite{lai2008} have studied both IRAC-detected and undetected
LAEs at $z=3.1$ by the stacking method, to find that LAEs posses
wide ranges of age ($\simeq 160$ Myr -- $1.6$ Gyr)
and mass ($\simeq 10^8$ -- $10^{10} M_\odot$)
without dust extinction.
\cite{finkelstein2008b} have analysed $14$ LAEs individually,
which are either IRAC-detected or undetected.
Their results also show wide ranges of
stellar population age ($\simeq 2.5$ -- $500$ Myr),
stellar mass ($\simeq 10^8$ -- $10^{10} M_\odot$),
and dust extinction ($A_{1200} \simeq 0.3$ -- $5.0$).

However,
most of the samples constructed to date are not necessarily large
and deep enough to constrain the average stellar populations of LAEs
and to study rare, massive LAEs which may be a bridge to
more massive and/or evolved objects like LBGs.
In addition, correlations between stellar population parameters have not been
addressed well.

Recently, \cite{ouchi2008} have constructed
the largest available sample of $z=3.1$ and $3.7$ LAEs
in an about $1$ deg$^2$
of the Subaru/\textit{XMM-Newton} Deep Field (SXDF)
from deep optical broadband and narrowband data.
These large survey data enable us
not only to find many massive LAEs,
but also to place better constraints
on less massive (i.e., average) LAEs using stacking analysis.
Our sample used in this study consists of
$224$ ($78$) LAEs at $z = 3.1$ ($3.7$)
covered by deep $JHK$ images
taken with the UKIRT/WFCAM
from UKIDSS Ultra Deep Survey \citep{warren2007}
and $3.6$ -- $8.0 \mu$m images
taken with the Spitzer/IRAC
from the Spitzer legacy survey of the UDS field
(SpUDS; PI: J. Dunlop)\footnotemark{}\footnotetext{\texttt{http://ssc.spitzer.caltech.edu/legacy/abs/dunlop.html}},
among which $5$ ($6$) are detected in the $K$ band
(i.e., brighter than the $3 \sigma$ detection limit
of the $K$-band image).
The $K$ band corresponds to the rest-frame $V$ and $B$ bands
at $z=3.1$ and $3.7$, respectively.
For $K$-undetected LAEs we make median stacked images
for individual bandpasses.
Since the vast majority ($96${\%}) are undetected in $K$,
the spectral energy distributions (SEDs)
constructed from stacking should closely represent
the whole LAE population at each redshift.
Since the time interval between the two redshifts is not so large
(the age of the universe is $\simeq 2.0$ Gyr at $z=3.1$
and $\simeq 1.7$ Gyr for $z=3.7$),
we treat the LAEs at $z=3.1$ and $3.7$ collectively as objects at $z \sim 3$,
without discussing possible evolution between the two redshifts.
In this paper, we show the results of
our stellar population analysis of these LAEs.
We also compare the stellar populations of LAEs with those
of other high redshift galaxy populations.
The contribution of LAEs to
the cosmic star formation density
and the stellar mass density is also evaluated.

The outline of this paper is as follows.
In Section 2, we present our data and LAE sample.
The SED fitting method is described in Section 3.
In Section 4, we present and discuss our SED fitting results.
A summary is given in Section 5.
Throughout this paper,
we use magnitudes in the AB system
and
assume a flat universe
with ($\Omega_m$, $\Omega_\Lambda$, $h$) $=$ ($0.3$, $0.7$, $0.7$).

\begin{table*}
\centering
\caption{Summary of Stellar Population Analysis of LAEs}
\label{tab:summary_of_earliers}
{\scriptsize
\begin{tabular}{cccccccc}
\hline
Reference
& Field
& Area
& redshift
& $N^\dagger$
& $m_{\rm threshold}^\ddagger$
& Bands$^\ast$
& Remark
\\

&
& [arcmin$^2$]
&
&
&  [mag]
&
&
\\
\hline
This Study
 &  SXDF
 &  $2340$
 &  $3.1$
 &  $5(200)$
 &  $K(3\sigma) = 24.1$
 &  $R, i, z, J, H, K, $[$3.6$], [$4.5$], [$5.8$], [$8.0$]
 &  1
\\
\cite{gawiser2006}
 &  ECDF-S
 &  $992$
 &  $3.1$
 &  $(18)$
 &  ---
 &  $U, B, V, R, I, z, J, K$
 &  2
 \\
\cite{gawiser2007}
 &  ECDF-S
 &  $992$
 &  $3.1$
 &  $(52)$
 &  $[3.6](2\sigma) = 25.2$
 &  $U, B, V, R, I, z, J, K, $[$3.6$], [$4.5$], [$5.8$], [$8.0$]
 &  3
 \\
\cite{lai2008}
 &  ECDF-S
 &  $992$
 &  $3.1$
 &  $18 (76)$
 &  $[3.6](2\sigma) = 25.2$
 &  $B, V, R, I, z, $[$3.6$], [$4.5$]
 &  4
 \\
\cite{nilsson2007}
 &  GOODS-S
 &  $44.4$
 &  $3.15$
 &  $(23)$
 &  $K_s(3\sigma) = 23.4$
 &  $U, B, V, i, z', J, H, K_s, $[$3.6$], [$4.5$], [$5.8$], [$8.0$]
 &  5
 \\
This Study
 &  SXDF
 &  $2340$
 &  $3.7$
 &  $6(61)$
 &  $K(3\sigma) = 24.1$
 &  $R, i, z, J, H, K, $[$3.6$], [$4.5$], [$5.8$], [$8.0$]
 &  1
\\
\cite{pentericci2008}
 &  GOODS-S
 &  $160$
 &  $3.5$ -- $6$
 &  $((38))$
 &  ---
 &  ---
 &  6
 \\
\cite{finkelstein2008b}
 &  GOODS-S
 &  $160$
 &  $4.5$
 &  $11 (3)$
 &  ---
 &  $NB, V, i', z', J, H, K, $[$3.6$], [$4.5$]
 &  7
 \\
\cite{pirzkal2007}
 &  HUDF
 &  $11$
 &  $4.00$ -- $5.76$
 &  $5 (4)$
 &  ---
 &  $z, J, H, K_s, $[$3.6$], [$4.5$]
 &  8
 \\
\cite{lai2007}
 &  GOODS-N
 &  $160$
 &  $5.7$
 &  $3$
 &  $[3.6](3\sigma) = 26.2$
 &  $i', z', $[$3.6$], [$4.5$]
 &  9
 \\
\cite{chary2005}
 &  Abell 370
 &  ---
 & $6.56$
 &  $1$
 &  ---
 &  ---
 &  10
 \\
\hline
\end{tabular}
} 

\medskip
\begin{minipage}{170mm}
\begin{flushleft}
$^\dagger$
Number of objects detected
in at least one rest-frame optical band.
The number in the parentheses is the number of objects
not detected in any of the rest-frame optical bands.

$^\ddagger$
Rest-frame optical detection threshold 
for LAEs, if given.

$^\ast$
Bands used for SED fitting, if given.

$^1$
$K$-detected objects are treated individually, while
$K$-undetected objects are stacked.

$^2$
All objects are stacked.

$^3$
All objects are stacked.

$^4$
IRAC-detected and IRAC-undetected objects are stacked separately.

$^5$
All objects are stacked.

$^6$
LBGs with EW(Ly$\alpha$) $>20$ {\AA},
and all are treated individually irrespective of the NIR
magnitude.

$^7$
All objects are treated individually irrespective of the NIR
magnitude.

$^8$
All objects are treated individually irrespective of the NIR
magnitude.

$^9$
All objects are treated individually.

$^{10}$
HCM 6A, a lensed LAE, detected at $3.6$ and $4.5 \mu$m.
SED fitting is performed considering
the effect of H$\alpha +$ [N\,{\sevensize II}] emission in the $4.5\mu$m band.

\end{flushleft}
\end{minipage}
\end{table*}


\section{DATA} \label{sec:data}

\subsection{Optical and NIR Images} \label{subsec:data}

Deep $BVRi'z'$ images of the SXDF were taken with Suprime-Cam
on the Subaru Telescope by the Subaru/\textit{XMM-Newton} Deep
Survey project \citep{furusawa2008}.
\cite{ouchi2008} combined this public data set with their own imaging
data taken with Suprime-Cam through two narrowband filters,
$NB503$ ($\lambda_c = 5029${\AA}, $\Delta\lambda = 74${\AA})
and
$NB570$ ($\lambda_c = 5703${\AA}, $\Delta\lambda = 69${\AA}),
and constructed samples of $368$ $z=3.1$ and $116$ $z=3.7$ LAEs
over a sky area of $\simeq 1$ deg$^2$.
These numbers are the logical sum of
the spectroscopically-confirmed and photometrically-selected LAEs
in \cite{ouchi2008}.
They are selected as objects with narrow-band excesses
and IGM absorption features shortward of Lyman $\alpha$
seen in color-color diagrams.
The line-of-sight depth corresponding to the FWHM of the
narrow-band filter is $55.7$ comoving Mpc for NB503
and $43.2$ comoving Mpc for NB570.
For all LAEs, $2''$-diametre aperture magnitudes
were measured in each bandpass \citep{ouchi2008}.


About $77${\%} of the SXDF Suprime-Cam field was imaged
in the $J$, $H$, and $K$ bands
with the wide-field near infrared camera WFCAM on the UKIRT
in the UKIDSS/UDS project \citep{lawlence2007}.
The UKIDSS/UDS is underway, and we use Data Release 3 for this study.
We align the $J,H,K$ images with the SXDS optical images
using common, bright stars, and then smooth them
with Gaussian filters so that the PSF sizes of the $J,H,K$ images
match those of the optical images (FWHM $\approx 1.''0$).
The $3\sigma$ limiting magnitudes over a $2''$-diametre aperture
are calculated to be $m_J = 24.5$, $m_H = 24.2$, and $m_K = 24.1$.
Because the zero-point magnitudes for the $J,H,K$ images are given
in the Vega system, we convert them into AB magnitudes using
the offset values given in Table 7 of \cite{hewett2006}.

\begin{figure}
 \begin{center}
  \includegraphics[scale=0.45]{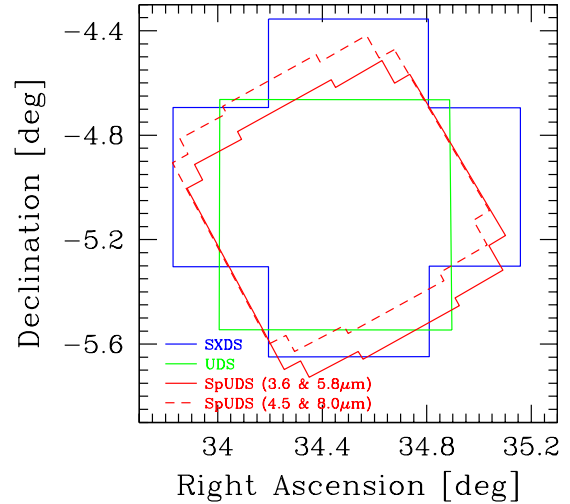}
 \end{center}
 \caption{
Observed field in the WCS coordinate system.
The blue cross-shaped region is the SXDF imaged with Suprime-Cam.
The green square corresponds to the UKIDSS/UDS (WFCAM) field.
The red solid line and red dashed line outline the SpUDS fields
with IRAC $3.6\mu$m and $5.8\mu$m imaging and $4.5\mu$m and
$8.0\mu$m imaging, respectively.
}
\label{fig:fields}
\end{figure}

The SpUDS covers $0.65$ deg$^2$ of the overlapping area
of the SXDS and UDS fields (Figure \ref{fig:fields}).
This $0.65$ deg$^2$ area corresponds to an effective survey volume
of
$4.6 \times 10^5$ Mpc$^3$ for $z=3.1$ LAEs
and
$4.1 \times 10^5$ Mpc$^3$ for $z=3.7$ LAEs,
respectively.
All of the SpUDS IRAC images are geometrically matched to the
optical images.
We calculate the $3\sigma$ limiting magnitude over a $3''$-diametre
aperture to be
$m_{3.6} = 24.8$, $m_{4.5} = 24.5$,
$m_{5.8} = 22.7$, and $m_{8.0} = 22.6$
in the $3.6$, $4.5$, $5.8$, and $8.0 \mu$m IRAC bands,
respectively.

\subsection{Photometry} \label{subsec:photo}

In this paper, we only analyse LAEs in the overlapping area of
$0.65$ deg$^2$ where the Suprime-Cam, WFCAM,
and IRAC data (either channel 1+3 or 2+4) are all available.
We perform $J$, $H$, and $K$ photometry with a $2''$-diametre aperture
at the position of LAEs in the narrowband images,
using the IRAF task \verb|apphot|.

We then convert $2''$-diametre aperture magnitudes in the optical
and $JHK$ bands into total magnitudes in the following manner.
First, we select $20$ bright and isolated point sources in the $R$-band
image, the deepest among all the images, and measure fluxes
in a $2''$ aperture and in a series of larger apertures
up to $5''$ with an interval of of $0.1''$.
Since we find the fluxes to level off for $> 4''$ apertures,
we define $4''$-aperture magnitudes as total magnitudes.
Then, we select $100$ point sources in the $R$-band image,
measure fluxes over $2''$ and $4''$ apertures,
and calculate an accurate offset between these two aperture magnitudes
to be $0.17$ mag.
For each of the optical and $JHK$ bandpasses, we subtract $0.17$ mag
from $2''$-aperture magnitudes to obtain total magnitudes for our LAEs.

For the Spitzer/IRAC four bands, we measure $3''$-diametre aperture
magnitudes for each LAE and converted them to total magnitudes
by applying the aperture correction given
by MUSYC survey\footnotemark{}\footnotetext{\texttt{http://www.astro.yale.edu/dokkum/SIMPLE/}\\\texttt{release\_1.1/doc/00README\_photometry}}.
The correction values are $0.52$, $0.55$, $0.74$, and $0.86$ mag
for $3.6\mu$m, $4.5\mu$m, $5.8\mu$m, and $8.0\mu$m, respectively.

\subsection{The LAE samples used for the stellar population analysis}

Table \ref{tab:summary_number} summarises our LAE samples
used for the stellar population analysis.
Our original sample consists of $224$ LAEs at $z=3.1$ and $78$ LAEs at $z=3.7$,
among which $35$ and $21$ have been spectroscopically confirmed,
including two newly confirmed objects
using Magellan/IMACS (Section \ref{subsec:spectroscopy}).

We regard an object as LAE with AGN and exclude it from our analysis
if it has a counterpart either
in the XMM-Newton X-ray catalogue \citep{ueda2008}
or in the VLA radio catalogue \citep{simpson2006},
or its spectrum has emission lines typical of AGN,
such as like SiIV, CIV, and HeII \cite[for details, see Section 4.6 in][]{ouchi2008}.
It should be noted, however, that we cannot completely exclude 
contamination from X-ray-faint, radio-quiet, and/or heavily obscured AGNs.

\begin{table*}
\centering
\caption{Summary of our samples used for the stellar population analysis}
\label{tab:summary_number}
\begin{tabular}{ccccc}
\hline
LAE sample
& $K$-band
& $N$
& with
& referred to as
\\

& detection
&
& spec-$z$
&
\\
\hline
$z=3.1$ & & & &
\\
\hline
Total
 &  --
 &  $224$
 &  $35$
 &  --
\\
LAE without AGN
 &  Yes
 &  $5$
 &  $5$
 &  $K$-detected LAEs
 \\
LAE with AGN
 &  Yes
 &  $2$
 &  $2$
 &  --
 \\
noise (false detection)
 &  Yes
 &  $1$
 &  $0$
 &  --
 \\
late-type star
 &  Yes
 &  $1$
 &  $0$
 &  --
 \\
LAE in the SpUDS fileds and without AGN
 &  No
 &  $200$
 &  $27$
 &  $K$-undetected LAEs
 \\
LAE out of the SpUDS fileds and/or with AGN
 &  No
 &  $15$
 &  $1$
 &  --
 \\
\hline
$z=3.7$ & & & &
\\
\hline
Total
 &  --
 &  $78$
 &  $21$
 &  --
\\
LAE without AGN
 &  Yes
 &  $6$
 &  $3$
 &  $K$-detected LAEs
 \\
LAE with AGN
 &  Yes
 &  $1$
 &  $1$
 &  --
 \\
confusion
 &  Yes
 &  $2$
 &  $2$
 &  --
 \\
LAE in the SpUDS fileds and without AGN
 &  No
 &  $61$
 &  $14$
 &  $K$-undetected LAEs
 \\
LAE out of the SpUDS fileds and/or with AGN
 &  No
 &  $8$
 &  $1$
 &  --
 \\
\hline
\end{tabular}
\end{table*}

\begin{table*}
 \caption{$K$-detected LAE sample}\label{tab:KdetectLAE_photo}
{\scriptsize
\begin{tabular}{ccccccccccccccc}
\hline
Object Name
  & $R$
  & $i'$
  & $z'$
  & $J$
  & $H$
  & $K$
  & $3.6 \mu$m
  & $4.5 \mu$m
  & $5.8 \mu$m
  & $8.0 \mu$m
  & $\log L($Ly$\alpha)$
  & EW(Ly$\alpha$)
  & $z$
  & Ref.
\\

  &
  &
  &
  &
  &
  &
  &
  &
  &
  &
  & [erg s$^{-1}$]
  & [{\AA}]
  &
  &
\\ \hline
NB503-N-21105 (R1)
 & 26.2 
 & 25.9 
 & 26.0 
 & 24.3 
 & 23.6 
 & 23.9 
 & 22.4 
 & 22.5 
 & 22.8$^\dagger$ 
 & 22.9$^\dagger$ 
 & 42.49 
 & 73 
 & 3.142
 & (1)
\\
NB503-N-42377
 & 23.8 
 & 23.8 
 & 23.9 
 & 23.8 
 & 23.8$^\dagger$ 
 & 23.0 
 & 23.7 
 & 24.2$^\dagger$ 
 & 23.1$^\dagger$ 
 & 22.3$^\dagger$ 
 & 43.50 
 & 77 
 & 3.154
 & (2)
\\
NB503-S-45244
 & 24.6 
 & 24.6 
 & 24.8 
 & 99.9$^\dagger$ 
 & 25.1$^\dagger$ 
 & 23.6 
 & 24.7$^\dagger$ 
 & ---
 & 99.9$^\dagger$ 
 & ---
 & 43.21 
 & 86 
 & 3.156
 & (2)
\\
NB503-S-65716
 & 23.9 
 & 23.9 
 & 24.1 
 & 24.3 
 & 23.8 
 & 23.5 
 & 24.1 
 & ---
 & 99.9$^\dagger$ 
 & ---
 & 43.38 
 & 65 
 & 3.114
 & (3)
\\
NB503-S-94275
 & 24.3 
 & 24.3 
 & 24.5 
 & 24.5$^\dagger$
 & 24.8$^\dagger$
 & 23.6 
 & 24.9$^\dagger$ 
 & 24.4$^\dagger$ 
 & 99.9$^\dagger$ 
 & 23.7$^\dagger$ 
 & 43.56 
 & 147 
 & 3.102
 & (2)
\\
NB570-N-32295 (R2)
 & 27.1 
 & 26.5 
 & 26.2 
 & 25.3$^\dagger$
 & 24.6$^\dagger$
 & 23.5 
 & 22.6 
 & 22.3 
 & 21.6 
 & 21.5 
 & 42.74 
 & 225 
 & 3.684
 & (1)
\\
NB570-S-84321
 & 24.9 
 & 24.8 
 & 25.0 
 & 24.4$^\dagger$
 & 24.0 
 & 23.7 
 & 24.3$^\dagger$ 
 & 23.7 
 & 26.7$^\dagger$ 
 & 24.2$^\dagger$ 
 & 42.99 
 & 55 
 & 3.648
 & (2)
\\
NB570-W-55371
 & 24.6 
 & 24.6 
 & 24.6 
 & 25.3$^\dagger$
 & 24.1$^\dagger$
 & 23.3 
 & 23.8 
 & 23.9 
 & 23.2$^\dagger$ 
 & 99.9$^\dagger$ 
 & 42.71 
 & 20 
 & 3.699
 & (2)
\\
NB570-C-24119
 & 25.5 
 & 25.2 
 & 25.1 
 & 26.8$^\dagger$
 & 24.7$^\dagger$
 & 23.2 
 & 23.0 
 & 22.5 
 & 22.0$^\dagger$ 
 & 22.0$^\dagger$ 
 & 42.58 
 & 34 
 & 3.69$^*$
 &
\\
NB570-S-88963 (R3)
 & 26.1 
 & 25.4 
 & 25.4 
 & 24.2 
 & 25.2$^\dagger$
 & 23.2 
 & 23.1 
 & 22.6 
 & 22.8$^\dagger$ 
 & 22.8$^\dagger$ 
 & 42.83 
 & 106 
 & 3.69$^*$
 &
\\
NB570-W-59558 (R4)
 & 25.7 
 & 25.4 
 & 25.2 
 & 24.2 
 & 24.0 
 & 23.0 
 & 23.3 
 & 23.1 
 & 99.9$^\dagger$ 
 & 23.8$^\dagger$ 
 & 43.12 
 & 150 
 & 3.69$^*$
 &
\\
\hline
\end{tabular}
} 

\medskip
\begin{minipage}{140mm}
\begin{flushleft}
\textbf{NOTES}: All magnitudes are total magnitudes.
No value means that the object is out of the image.
Negative fluxes have been replaced with $99.9$ mag.
Sources of spectroscopic redshifts are
(1) this study, (2) \cite{ouchi2008}, (3) Ouchi et al. in preparation.

$^\dagger$Fainter than the $3\sigma$ limiting magnitude.

$^*$Not spectroscopically confirmed.
we give $z=3.69$, which corresponds to the central wavelength of NB570.
\end{flushleft}
\end{minipage}
\end{table*}

In the stellar population analysis, we treat objects detected in
the $K$ band individually, while we stack $K$-undetected objects
at each redshift to make an average SED and fit it with stellar
population synthesis models.
The $K$ band corresponds to the rest-frame $V$ and $B$ bands
for $z=3.1$ and $z=3.7$, respectively, both of which are redward
of the $4000$ {\AA} break.
We find that the stacked objects of the two redshifts are both
detected in $K$.
Since as high as $96\%$ of our sample are undetected in $K$
(see below), the stacked SEDs of $K$-undetected objects
is considered to represent closely the whole LAE population at the two redshifts.

\subsection{$K$-detected LAEs} \label{subsec:detected}

Among a total of $224$ ($78$) LAEs at $z=3.1$ ($3.7$),
$9$ ($9$) are found to be brighter than
the $K$-band $3 \sigma$ magnitude (i.e., $24.1$ mag).
Among these $18$,
$2$ ($1$) objects at $z=3.1$ ($3.7$) are found to host AGN 
(judged from the X-ray and radio data and spectroscopic data 
as described above),
and $2$ objects at $z=3.7$ are found to be significantly confused
by their neighbouring objects.
we do not use them for the stellar population analysis.
We then visually inspect the images of the remaining 
$13$ $K$-bright objects,
and find that one object is likely to be a false detection,
because it is not detected in any of the other IR bands
($J$, $H$, and IRAC bands) and it falls on a low quality part of
the $K$-band image.
Next, we examine the SEDs of the remaining $12$ objects over the range
from the $B$ band to the IRAC $8.0\mu$m band,
to find that one has an SED consistent with a late-type star,
with a peak near the $K$ band.
After removal of these two objects, the number of LAEs
brighter than $K(3\sigma)$ is
five for $z=3.1$ and six for $z=3.7$.
Hereafter, we call them $''K$-detected LAEs$''$.
Table \ref{tab:KdetectLAE_photo}
summarises their photometry, $L({\rm Ly}\alpha)$,
EW$({\rm Ly}\alpha)$, and redshift.

We have found that the source center of NB503-N-21105 in the narrow band (Ly$\alpha$)
offsets from those in the broad bands (continuum) by about one arcsec.
We thus cannot rule out the possibility that
NB503-N-21105 is an LAE with a chance projection of a galaxy along the line-of-sight.
However, the two components are more likely to be physically connected,
because the narrow-band image is elongated to the direction of the offset.

NB570-N-32295 has a power-law like spectrum over the whole
wavelength range up to $8.0 \mu$m.
Although we find that this spectrum is reproduced reasonably
well by a stellar system (see Section \ref{subsec:SEDfitting-results}),
we cannot completely rule out a possible contamination from an AGN.

\subsection{New Spectroscopy of Two Red Objects}
\label{subsec:spectroscopy}

Six out of the $11$ $K$-detected LAEs have been
spectroscopically identified by \cite{ouchi2008} and Ouchi et al. in preparation.
Among the remaining five, we selected two red objects,
NB503-N-21105 and NB570-N-32295, and carried out spectroscopy
with the Inamori Magellan Areal Camera and Spectrograph
(IMACS) on the Magellan I Baade $6.5$m telescope
at the Las Campanas Observatory.
The observation was made on 2008 December 19
with the WB$4800$ -- $7800$ filter
and the $300$ lines mm$^{-1}$ grism
whose blaze angle is $17.5$ degrees.
We chose a slit width of $1.0''$.
The on-source exposure time was
$15300$ seconds under the
$0.5''$ -- $0.75''$ seeing condition.
The spectral coverage was $4800$ -- $7800$ {\AA}.
The spectral resolution and the corresponding velocity resolution
are $R \simeq 890$ and $\Delta v \simeq 340$ km s$^{-1}$ at $5029$ {\AA},
and $\simeq 1000$ and $300$ km s$^{-1}$ at $5703$ {\AA}.
The data were reduced with the COSMOS pipeline
version 2.12\footnote{\texttt{http://www.ociw.edu/Code/cosmos}}.

\begin{figure*}
 \begin{center}
  \includegraphics[scale=0.85]{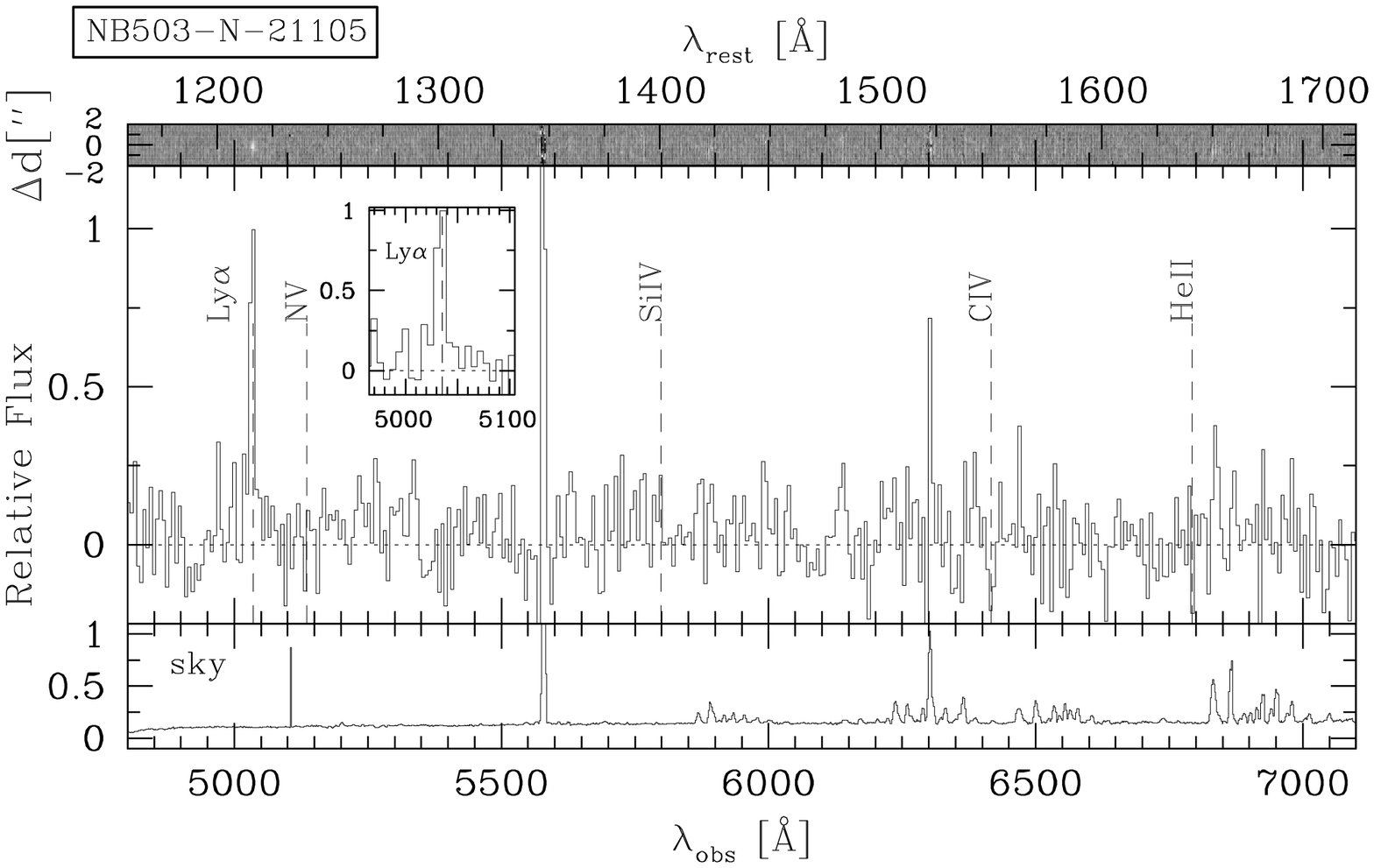}
  \includegraphics[scale=0.85]{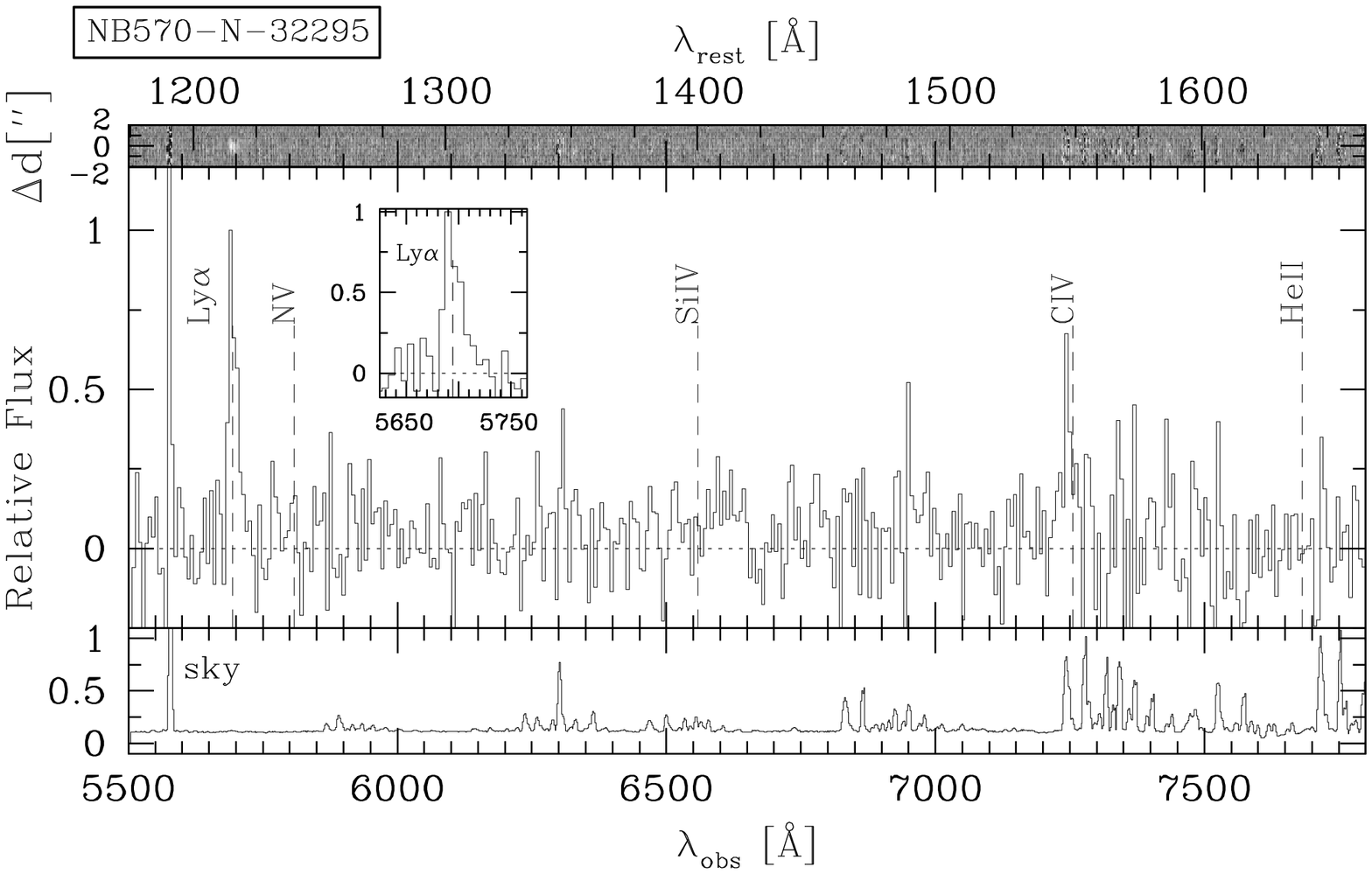}
\end{center}
 \caption{
Spectra of NB503-N-21105 (top) and NB570-N-32295 (bottom).
For each object, the top and middle panels show
the two- and one-dimensional spectra, respectively.
The one-dimensional spectra have been
smoothed with a $3$ pixel boxcar.
The dashed lines with the legend
correspond to the wavelengths
of typical emission lines from AGN.
Inserted is a zoom up around Ly$\alpha$ line.
The bottom panel shows the sky background with an arbitrary normalisation.
}
\label{fig:2d1dspec}
\end{figure*}

We present the reduced 2D and 1D spectra in Figure \ref{fig:2d1dspec}.
Both objects have a single, strong emission line
($> 6\sigma$ detection for NB503-N-21105
and $> 4\sigma$ detection for NB570-N-32295)
and have no other emission line or detectable continuum 
emission\footnote{There 
is an emission-like feature around [C\,{\sevensize IV}]
in the spectrum of NB570-N-32295.
We find, however, that there is a strong sky emission line at
this wavelength and that this feature is statistically insignificant, 
being consistent with a residual of sky subtraction.}.
The central wavelength of the single line is
$5035.4$ {\AA} for NB503-N-21105,
and $5693.9$ {\AA} for NB570-N-32295.
We conclude that both objects are a real LAE from the following discussion.
If NB503-N-21105 is an [O\,{\sevensize III}] ([O\,{\sevensize II}]) emitter,
H$\alpha$ ([O\,{\sevensize III}]) emission should be detected
at $\simeq 6629$ {\AA} ($6759$ {\AA}).
Similarly, if NB570-N-32295 is an [O\,{\sevensize III}]
([O\,{\sevensize II}]) emitter,
H$\alpha$ ([O\,{\sevensize III}]) emission should be seen
at $\simeq 7482$ {\AA} ($7661$ {\AA}).
We summarise the spectroscopic results
in Table \ref{tab:RedLAE_specresults}.

\begin{table*}
\centering
 \caption{Spectroscopic results of two red LAEs}\label{tab:RedLAE_specresults}
\begin{tabular}{ccccc}
\hline
Object Name
  & RA (J2000)
  & Dec (J2000)
  & FWHM(Ly$\alpha$)$^\dagger$
  & $z$
\\

  & [h:m:s]
  & [d:m:s]
  & [km s$^{-1}$]
  &
\\ \hline
NB503-N-21105 (R1)
 & 2:18:42.186
 & $-$4:46:38.54
 & $205 \pm 129$
 & 3.142 \\
NB570-N-32295 (R2)
 & 2:17:25.630
 & $-$4:44:33.57
 & $629 \pm 201$
 & 3.684 \\
\hline
\end{tabular}

\medskip
\begin{minipage}{140mm}
\begin{flushleft}
$^\dagger$After correction
for the instrumental broadening
on the assumption of a Gaussian profile.
\end{flushleft}
\end{minipage}
\end{table*}


\subsection{Stacking of $K$-undetected LAEs} \label{subsec:stacking}

\begin{table*}
\caption{Stacked median LAE sample.}\label{tab:KundetectLAE_photo}
{\scriptsize
\begin{tabular}{cccccccccccccc}
\hline
Object Name
  & $R$
  & $i'$
  & $z'$
  & $J$
  & $H$
  & $K$
  & $3.6 \mu$m
  & $4.5 \mu$m
  & $5.8 \mu$m
  & $8.0 \mu$m
  & $\log L($Ly$\alpha)$
  & EW(Ly$\alpha$)
  & $\langle z\rangle$
\\

  &
  &
  &
  &
  &
  &
  &
  &
  &
  &
  & [erg s$^{-1}$]
  & [{\AA}]
  &
\\ \hline
NB503-$K$-undetected
 & 26.9 
 & 26.9 
 & 27.1 
 & 27.4$^\dagger$ 
 & 28.8$^\dagger$ 
 & 26.5 
 & 27.1 
 & 27.4$^\dagger$ 
 & 99.9$^\dagger$ 
 & 99.9$^\dagger$ 
 & 42.54 
 & 155 
 & 3.14$^*$ \\
NB570-$K$-undetected
 & 26.4 
 & 26.3 
 & 26.4 
 & 26.9$^\dagger$ 
 & 27.2$^\dagger$ 
 & 25.4 
 & 26.5 
 & 29.2$^\dagger$ 
 & 25.5$^\dagger$ 
 & 99.9$^\dagger$ 
 & 42.82 
 & 135 
 & 3.69$^*$ \\
\hline
\end{tabular}
} 

\medskip
\begin{minipage}{140mm}
\begin{flushleft}
\textbf{NOTES}: All magnitudes are total magnitudes.
Negative fluxes have been replaced with $99.9$ mag.

$^\dagger$Fainter than the $3\sigma$ limiting magnitude.

$^*$Redshifts corresponding to the central wavelengths of
NB503 and NB570.
\end{flushleft}
\end{minipage}
\end{table*}

\begin{figure}
 \begin{center}
  \includegraphics[scale=0.7]{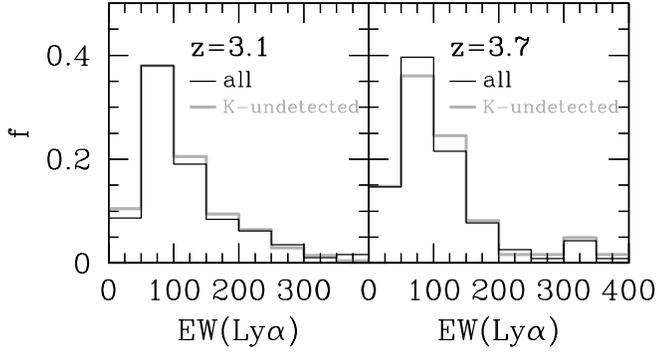}
\end{center}
 \caption{
Normalised histograms of the rest-frame Ly$\alpha$ EWs for
all (black) and $K$-undetected (grey) LAEs
at $z=3.1$ (left) and $3.7$ (right).
}
\label{fig:histo_EW}
\end{figure}

Among the LAEs fainter than the $K(3\sigma)$ magnitude,
$200$ ($61$) at $z=3.1$ ($3.7$) have
all $10$ broadband data from $R$ to $8.0 \mu$m
and do not show AGN features, 
to be used for the stellar population analysis.
We call these LAEs $''K$-undetected LAEs$''$.
We make median-stacked multi-waveband images
separately for the two redshifts.
Table \ref{tab:KundetectLAE_photo}
summarises their photometry, average $L({\rm Ly}\alpha)$,
EW$({\rm Ly}\alpha)$ derived from narrow-band
and $R$-band magnitudes, and redshift.
In Figure \ref{fig:histo_EW}, we show the histograms
of the rest-frame Ly$\alpha$ EWs for all and $K$-undetected LAEs,
to confirm that the $K$-undetected LAEs are representative
of the whole sample.

\subsection{Applications of Broadband Color Selections}

\subsubsection{LBG Selection}

We examine whether the six $K$-detected LAEs
at $z=3.7$ are selected as LBGs,
by applying the $BRi'$ colour selection criteria
for $z \sim 4$ LBGs \citep[e.g.][]{ouchi2004,yoshida2006}.
None of them are found to meet the criteria.
This is probably because our objects tend to have
redder UV slopes than typical LBGs,
or the $B$-band image is not deep
enough to detect Lyman break,
or the redshift of $z=3.7$ is
close to the edge of the redshift range
covered by the $BRi'$ selection criteria.

In addition, we apply the $BRi'$ colour criteria
to the $61$ $K$-undetected LAEs at $z=3.7$
and find that only seven satisfy the criteria.
We also find that the stacked LAE at $z=3.7$ does not meet
the criteria.

\subsubsection{DRG Selection}

We then apply the colour criterion for distant-red galaxies (DRGs),
$J-K>1.3$, to our sample, and find that
none of the $K$-detected LAEs and the two stacked LAEs
satisfies the DRG criterion.
We also apply the colour criterion to
the best-fit model SEDs for these LAEs
derived from our SED fitting (Section \ref{sec:SEDfitting}).
Among the 11 $K$-detected LAEs, model SEDs for 
three (NB503-N-21105, NB570-N-32295,
and NB570-C-24119) meet the criterion.
The former two have very red SEDs
and have very large dust extinction
($E(B-V)_\star \sim 0.65$).
We will focus on this kind of LAEs
in Section \ref{subsec:red-laes}.


\section{SED FITTING} \label{sec:SEDfitting}

\subsection{Method} \label{subsec:SEDfitting-method}

After obtaining rest-frame UV-to-optical photometry for the
$K$-detected and stacked LAEs, we analyse their stellar populations
by the standard SED fitting method \citep[e.g.,][]{furusawa2000,papovich2001}.
We use the stellar population synthesis model of GALAXEV
\citep[][hereafter BC03]{bc2003} to produce model SEDs.
Most of the previous studies have used BC03 
(Table \ref{tab:sum_of_assumption}).
We make a large set of mass-normalised model SEDs, 
varying star-formation timescale, age,
and dust extinction (or $E(B-V)_\star$).
These SEDs are then redshifted to $z=3.1$ and $z=3.7$
and convolved with ten bandpasses, $R,i,z,J,H,K$, and IRAC four bands,
to calculate flux densities.
For each LAE, we search for the best-fit SED that minimizes
\begin{equation}
\chi^2
       = \sum_i \frac{\left[ f^{(i)}_{\rm obj} - M_{\rm star}f^{(i)}_{\rm model} \left({\rm age}, E(B-V)_\star, \tau \right) \right]^2}{\sigma^2 ( f^{(i)}_{\rm obj} )},
\label{eq:chi2}
\end{equation}
where
$f^{(i)}_{\rm obj}$ is the observed flux density in the $i$-th bandpass,
$f^{(i)}_{\rm model}$ is the mass-normalised model flux density in the $i$-th bandpass,
$\tau$ is the star-formation timescale,
and $\sigma ( f^{(i)}_{\rm obj} )$ is the photometric error
in the $i$-th bandpass.
Because $M_{\rm star}$ is the amplitude of a model SED,
we obtain the best-fit $M_{\rm star}$ for each set of 
(age, $E(B-V)_\star$, $\tau$) 
by solving $\partial \chi^2 / \partial M_{\rm star} = 0$, 
and calculate $\chi^2$. 
Then we search for the set of the best-fit parameters
that gives the minimum $\chi^2$.
The errors in the best-fit SED parameters  
correspond to $1 \sigma$ confidence interval ($\Delta \chi^2 < 1$) for each parameter.

\begin{table*}
\centering
\caption{Summary of
Stellar Population Analysis of LAEs:
Assumption on SED Modeling}
 \label{tab:sum_of_assumption}
{\scriptsize
\begin{tabular}{ccccccc}
\hline
Reference
& Model$^{\dagger}$
& IMF
& SFH$^{\ddagger}$
& Metallicity
& Extinction$^{\ast}$
& Remarks
\\

&
&
&
& [$Z_\odot$]
& Curve
&
\\
\hline
This Study
 &  BC03, CB08
 &  Salpeter
 &  const, exp
 &  $0.005$, $0.02$, $0.2$, $1$
 &  C00
 &  (O09)
\\
\cite{gawiser2006}
 &  BC03
 &  Salpeter
 &  const
 &  $1$
 &  C97
 &
\\
\cite{gawiser2007}
 &  M05
 &  Salpeter
 &  two-burst
 &  $0.02$ -- $1$
 &  C00
 &
\\
\cite{lai2008}
 &  BC03
 &  Salpeter
 &  const
 &  $1$
 &  C00
 &
\\
\cite{nilsson2007}
 &  BC03
 &  Salpeter
 &  const
 &  $0.005$, $0.2$, $1$
 &  CF00
 &
\\
\cite{pentericci2008}
 &  BC03, M05, CB08
 &  Salpeter
 &  exp
 &  $0.02$, $0.2$, $1$, $2.5$
 &  C00
 &  (P09)
\\
\cite{finkelstein2008b}
 &  BC03
 &  Salpeter
 &  exp, two bursts
 &  $0.005$, $0.02$, $0.2$, $0.4$, $1$
 &  C94
 &  (F09)
\\
\cite{pirzkal2007}
 &  BC03
 &  Salpeter
 &  ssp, exp, 2bp
 &  $0.005$ -- $2.5$
 &  C00
 &
\\
\cite{lai2007}
 &  BC03
 &  Salpeter
 &  ssp, const
 &  $0.005$, $1$
 &  C00
 &
\\
\cite{chary2005}
 &  BC03
 &  Salpeter
 &  ---
 &  $0.02$
 &  ---
 &
\\
\hline
\end{tabular}
} 
\medskip
\begin{minipage}{165mm}
\begin{flushleft}

$^{\dagger}$
Population synthesis model.
$''$BC03$''$, $''$M05$''$, and $''$CB08$''$ represent
\cite{bc2003}, \cite{maraston2005},
and the unpublished Charlot {\&} Bruzual model, respectively.

$^{\ddagger}$
Star formation history.
$''$ssp$''$: instantaneous starburst,
$''$const$''$: constant star formation history,
$''$exp$''$: exponentially decaying star formation
history (and $\tau$ is $e$-folding time),
$''$two-burst$''$: superposition of an old instantaneous burst component
and a young $''$exp$''$ component,
$''$two bursts$''$: superposition of an old "exp" component (age $= 1.4$ Gyr and $\tau = 10^5$ yr)
and a young $''$exp$''$ component ($\tau = 10^5$ yr),
$''$2bp$''$: superposition of two different instantaneous starbursts.
The $e$-folding times examined are:
$\tau = 1$, $10$, $10^2$, $10^3$ Myr \textbf{(O09)},
$\tau = 0.1$, $0.3$, $0.6$, $1$, $1.5$, $2$, $3$, $4$, $5$, $7$, $9$,
$13$, $15$ Gyr \textbf{(P09)},
$\tau = 10^5$, $10^6$, $10^7$, $10^8$, $4 \times 10^9$yr
\textbf{(F09)}.

$^{\ast}$
Dust extinction law.
C00: \cite{calzetti2000},
C97: \cite{calzetti1997},
C94: \cite{calzetti1994},
CF00: \cite{charlot2000}.

\end{flushleft}
\end{minipage}
\end{table*}

We do not use either $B$- or $V$-band photometry in the fitting,
since $B$-band photometry suffers from the IGM absorption shortward
of the Lyman $\alpha$ wavelength and $V$-band photometry is
contaminated from Ly$\alpha$ emission.
The amount of the IGM absorption considerably differs by the line of sight. 
Ly$\alpha$ fluxes can be estimated from narrow and broad band photometry,
but they have an uncertainty of a factor of $\sim 2$ \citep[see Figure 15 of][]{ouchi2008}.
The shortest broadband we use for SED fitting is the $R$ band,
which corresponds to $1585$ {\AA} and $1380$ {\AA} for $z=3.1$ and $3.7$ LAEs.
These wavelengths are very close to Ly$\alpha$ and short enough to cover the wide range of SEDs.

We adopt Salpeter's initial mass function \citep{salpeter1955}
with lower and upper mass cutoffs of $0.1$ and $100 M_\odot$.
We fix the metallicity to $Z/Z_\odot = 0.2$, considering the
fact that LBGs at $z \sim 3$ tend to have subsolar metallicies
at $Z / Z_\odot \sim (1/4)$ -- $(1/3)$
\cite[a strongly lensed LBG, cB58:][]{pettini2000,teplitz2000}
and
$Z/Z_\odot \sim (1/10) $ -- $(2/3)$
\cite[four bright LBGs:][]{pettini2001}.
Although \cite{shapley2004} reported that
very massive $z \sim 2$ LBGs with $M_{\rm star} \sim 10^{11} M_\odot$
have approximately the solar metallicity,
\cite{erb2006} found that
less massive ($\la 10^{10} M_\odot$) LBGs at similar redshifts
have subsolar metallicities
\citep[see also,][]{maiolino2008}.
It appears to be reasonable that the metallicites of typical
LAEs are lower than (and at most comparable to) those of
typical LBGs.

We also examine models with $Z/Z_\odot = 0.005$ and $1$,
and find that while the best-fit stellar mass is insensitive to
metallicity over this range, the best-fit age and dust extinction
become higher with decreasing metallicity.
However, the dependencies of age and dust extinction on metallicity
are not so strong:
for both parameters, the change from the best-fit value
for the $Z/Z_\odot = 0.02$ model is within its $1\sigma$ errors.

The star formation timescales examined are
(i) constant star formation
and (ii) exponentially decaying star formation
with four $e$-folding times of
$\tau = 0.001, \, 0.01, \, 0.1, \, 1$ Gyr,
and we search for the best-fit SED separately for
the cases (i) and (ii).
As explained in the next section,
we adopt the results for constant star formation
for our discussion of stellar populations.
For dust extinction, we use Calzetti's extinction law
\citep{calzetti2000} and vary $E(B-V)_\star$
over $0$ and $1.50$ with an interval of $0.01$.

Models of constant star formation have
three free parameters,
stellar mass, age, and dust extinction,
while models of exponentially decaying star formation
have one more parameter, $e$-folding time.

\subsection{Results} \label{subsec:SEDfitting-results}

\begin{figure*}
 \begin{center}
  \includegraphics[scale=0.92]{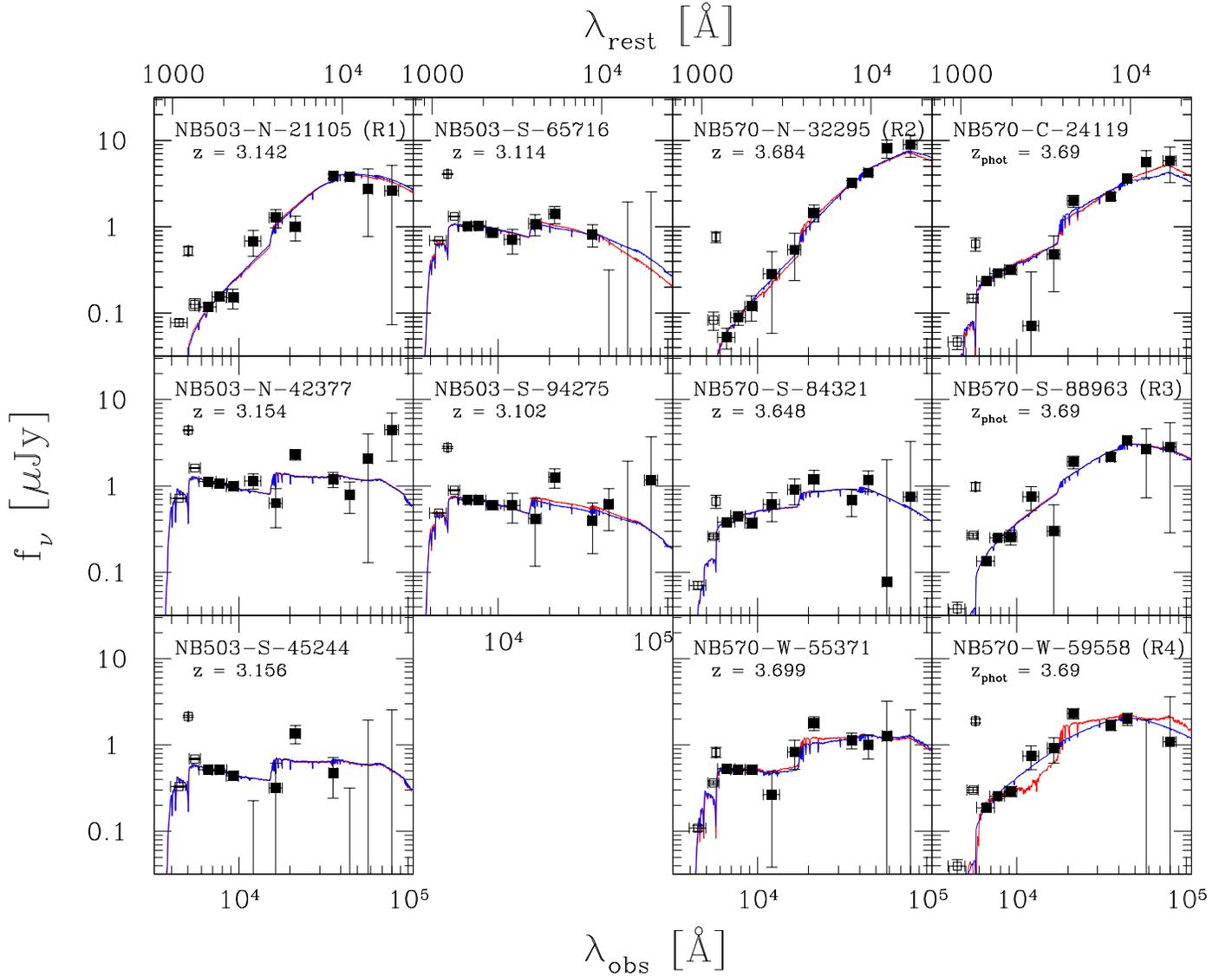}
\end{center}
 \caption[The observed photometries with the best-fit model SED
of the $K$-detected LAEs.]
{
Best-fit model SEDs (curves) and observed photometry (squares)
for $K$-detected LAEs.
The blue and red curves
correspond to constant star formation
and exponentially decaying star formation, respectively.
Data shown by open squares are not used for the SED fitting.
}
\label{fig:SEDs}
\end{figure*}

\begin{figure*}
 \begin{center}
  \includegraphics[scale=1.2]{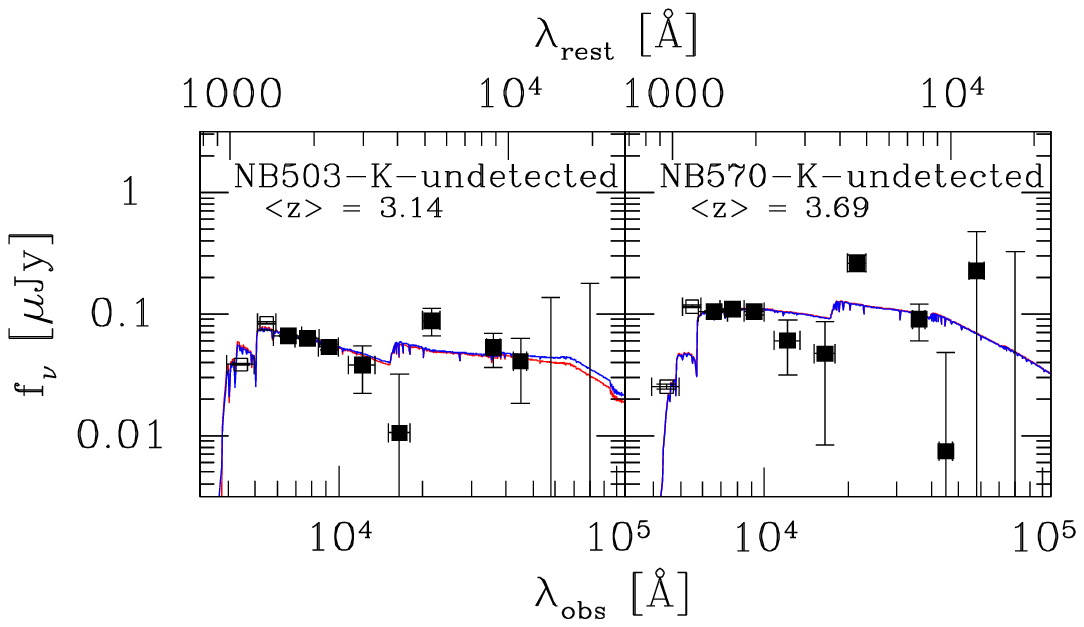}
\end{center}
 \caption[The observed photometries with the best-fit model SED
when constant star formation is assumed.]
{
Same as Figure \ref{fig:SEDs},
but for $K$-undetected (stacked) LAEs.
}
\label{fig:SEDs2}
\end{figure*}

Figures \ref{fig:SEDs} and \ref{fig:SEDs2} show the results
of the SED fitting for objects at $z=3.1$ and $3.7$.
For each object, the blue and red curves correspond respectively
to the best-fit SEDs for constant star formation and
for exponentially decaying star formation.
Both SEDs give similarly good fits
for most of the objects,
implying a difficulty in
constraining the star formation history.

The best-fit parameters are summarised in
Table \ref{tab:SEDfitting} for constant star formation
and Table \ref{tab:SEDfitting_tau} for exponentially decaying
star formation.
The reduced $\chi^2$ tends to be slightly larger
for the model of exponentially decaying star formation.
This may be because the $e$-folding time interval adopted is
too large ($1$ dex) to fine-tune model SEDs.
In any case, adding the $e$-folding time as a free parameter
does not significantly improve the fit.

\begin{table*}
\begin{minipage}{140mm}
\caption{SED Fitting Result ($Z = 0.2 Z_\odot$, constant SFH)}\label{tab:SEDfitting}
\begin{tabular}{cccccc}
\hline
Object Name
  & $\log M_{\rm star}$
  & $E(B-V)_\star$
  & $\log$(Age)
  & $\log $(SFR)
  & $\chi^2_r$ $^*$
\\
  & $[M_\odot]$
  & [mag]
  & [yr]
  & [$M_\odot$ yr$^{-1}$]
  &
\\ \hline
& & $z=3.1$
& & & \\ \hline
   NB503-N-21105 (R1)
 & $10.43^{+0.22}_{-0.13}$
 &  $0.70^{+0.05}_{-0.05}$
 &  $6.68^{+0.28}_{-0.12}$
 &  $3.75^{+0.34}_{-0.40}$
 &  2.36 \\
   NB503-N-42377
 & $9.82^{+0.09}_{-0.10}$
 &  $0.00^{+0.02}_{-0.00}$
 &  $8.51^{+0.10}_{-0.20}$
 &  $1.40^{+0.09}_{-0.01}$
 &  2.75 \\
   NB503-S-45244
 & $9.58^{+0.23}_{-0.34}$
 &  $0.00^{+0.07}_{-0.00}$
 &  $8.61^{+0.25}_{-0.70}$
 &  $1.06^{+0.36}_{-0.01}$
 &  2.17 \\
   NB503-S-65716
 & $9.19^{+0.13}_{-0.08}$
 &  $0.13^{+0.05}_{-0.03}$
 &  $6.98^{+0.40}_{-0.34}$
 &  $2.22^{+0.46}_{-0.26}$
 &  1.91 \\
   NB503-S-94275
 & $8.97^{+0.58}_{-0.13}$
 &  $0.11^{+0.07}_{-0.11}$
 &  $7.10^{+1.26}_{-0.48}$
 &  $1.88^{+0.59}_{-0.68}$
 &  0.96 \\ 
  NB503-$K$-undetected
 & $8.12^{+0.26}_{-0.37}$
 &  $0.03^{+0.05}_{-0.03}$
 &  $7.81^{+0.45}_{-0.81}$
 &  $0.36^{+0.44}_{-0.17}$
 &  1.39  \\ 
\hline 
& & $z=3.7$
& & & \\ \hline
   NB570-N-32295 (R2)
 & $10.60^{+0.54}_{-0.11}$
 &  $0.67^{+0.04}_{-0.25}$
 &  $7.40^{+1.81}_{-0.38}$
 &  $3.23^{+0.27}_{-1.19}$
 &  0.57 \\
   NB570-S-84321
 & $9.59^{+0.23}_{-0.16}$
 &  $0.32^{+0.05}_{-0.08}$
 &  $6.68^{+0.85}_{-0.12}$
 &  $2.91^{+0.35}_{-0.73}$
 &  1.34 \\
   NB570-W-55371
 & $10.19^{+0.10}_{-0.17}$
 &  $0.04^{+0.04}_{-0.04}$
 &  $8.96^{+0.25}_{-0.35}$
 &  $1.33^{+0.19}_{-0.15}$
 &  1.15 \\
   NB570-C-24119
 & $10.67^{+0.17}_{-0.15}$
 &  $0.29^{+0.07}_{-0.08}$
 &  $8.71^{+0.50}_{-0.50}$
 &  $2.06^{+0.33}_{-0.33}$
 &  1.39 \\
   NB570-S-88963 (R3)
 & $10.54^{+0.29}_{-0.33}$
 &  $0.62^{+0.06}_{-0.10}$
 &  $6.56^{+0.44}_{-0.08}$
 &  $3.98^{+0.37}_{-0.75}$
 &  1.93 \\
   NB570-W-59558 (R4)
 & $10.71^{+0.03}_{-0.69}$
 &  $0.63^{+0.01}_{-0.16}$
 &  $6.16^{+0.74}_{-1.06}$
 &  $4.55^{+1.09}_{-1.42}$
 &  2.14 \\
   NB570-$K$-undetected
 & $8.50^{+0.20}_{-0.10}$
 &  $0.19^{+0.04}_{-0.03}$
 &  $6.76^{+0.16}_{-0.22}$
 &  $1.74^{+0.42}_{-0.22}$
 &  3.27 \\
\hline
\end{tabular}

\medskip
\begin{flushleft}
$^*$Reduced $\chi$ squares.
The degree of freedom is $5 \, (= 8 - 3)$
for NB503-S-45244 and NB503-S-65716,
and $7 \, (= 10 - 3)$ for the others.
\end{flushleft}
\end{minipage}
\end{table*}

\begin{table*}
\begin{minipage}{140mm}
\caption{SED Fitting Result ($Z = 0.2 Z_\odot$, exponentially decaying SFH) }\label{tab:SEDfitting_tau}
\begin{tabular}{ccccccc}
\hline
Object Name
  & $\log M_{\rm star}$
  & $E(B-V)_\star$
  & $\log$(Age)
  & $\log$(SFR)
  & $e$-folding time
  & $\chi^2_r$ $^*$
\\

  & $[M_\odot]$
  & [mag]
  & [yr]
  & [$M_\odot$ yr$^{-1}$]
  & [Gyr]
  &
\\ \hline
& & & $z=3.1$
& & \\ \hline
 NB503-N-21105 (R1)
 & $10.13^{+0.51}_{-0.02}$
 &  $0.64^{+0.11}_{-0.33}$
 &  $6.64^{+1.52}_{-0.12}$
 &  $2.24^{+1.79}_{-\infty}$
 &  0.001
 &  2.71 \\
 NB503-N-42377
 & $9.82^{+0.05}_{-0.12}$
 &  $0.00^{+0.02}_{-0.00}$
 &  $8.46^{+0.05}_{-0.35}$
 &  $1.38^{+0.09}_{-0.10}$
 &  1
 &  3.21 \\
 NB503-S-45244
 & $9.54^{+0.21}_{-0.81}$
 &  $0.00^{+0.07}_{-0.00}$
 &  $8.51^{+0.20}_{-1.57}$
 &  $1.05^{+0.36}_{-2.54}$
 & 1
 &  2.72 \\
 NB503-S-65716
 & $8.99^{+0.12}_{-0.04}$
 &  $0.11^{+0.03}_{-0.04}$
 &  $6.66^{+0.26}_{-0.04}$
 &  $1.01^{+0.26}_{-1.53}$
 & 0.001
 &  2.10 \\
 NB503-S-94275
 & $8.87^{+0.62}_{-0.19}$
 &  $0.06^{+0.10}_{-0.05}$
 &  $6.92^{+1.29}_{-0.38}$
 &  $-0.73^{+3.11}_{-1.66}$
 & 0.001
 &  1.05 \\
 NB503-$K$-undetected
 & $7.75^{+0.61}_{-0.12}$
 &  $0.01^{+0.05}_{-0.01}$
 &  $7.00^{+1.21}_{-0.10}$
 &  $-2.57^{+3.10}_{-0.80}$
 &  0.001
&  1.56 \\ 
\hline 
& & &
$z=3.7$ & & \\ \hline
 NB570-N-32295 (R2)
 & $10.84^{+0.14}_{-0.37}$
 &  $0.48^{+0.22}_{-0.05}$
 &  $7.81^{+0.85}_{-0.73}$
 &  $-22.91^{+26.29}_{-\infty}$
 &  0.001
 &  0.56 \\
 NB570-S-84321
 & $9.54^{+0.42}_{-0.30}$
 &  $0.31^{+0.06}_{-0.30}$
 &  $6.68^{+1.33}_{-0.16}$
 &  $2.75^{+0.51}_{-\infty}$
 &  0.01
 &  1.55 \\
 NB570-W-55371
 & $10.05^{+0.04}_{-0.07}$
 &  $0.00^{+0.03}_{-0.00}$
 &  $8.46^{+0.00}_{-0.05}$
 &  $0.93^{+0.14}_{-0.00}$
 & 0.1
 &  1.08 \\
  NB570-C-24119
 & $10.39^{+0.37}_{-0.07}$
 &  $0.29^{+0.05}_{-0.06}$
 &  $7.60^{+1.30}_{-0.16}$
 &  $-12.92^{+15.17}_{-6.41}$
 & 0.001
 &  1.53 \\
 NB570-S-88963 (R3)
 & $10.54^{+0.25}_{-0.39}$
 &  $0.63^{+0.05}_{-0.10}$
 &  $6.50^{+0.44}_{-0.04}$
 &  $3.18^{+1.09}_{-0.76}$
 &  0.001
 &  2.23 \\
 NB570-W-59558 (R4)
 & $10.46^{+0.02}_{-0.08}$
 &  $0.00^{+0.01}_{-0.00}$
 &  $8.56^{+0.00}_{-0.05}$
 &  $-12.08^{+1.65}_{-\infty}$
 & 0.01
 &  1.85 \\
 NB570-$K$-undetected
 & $8.49^{+0.20}_{-0.26}$
 &  $0.19^{+0.04}_{-0.06}$
 &  $6.74^{+0.18}_{-0.22}$
 &  $1.74^{+0.41}_{-2.29}$
 & 0.1
 &  3.82
\\
\hline
\end{tabular}

\medskip
\begin{flushleft}

$^*$Reduced $\chi$ squares.
The degree of freedom is
$4 \, (= 8 - 4)$
for NB503-S-45244 and NB503-S-65716,
and $6 \, (= 10 - 4)$ for the others.
\end{flushleft}
\end{minipage}
\end{table*}

It is found from these tables
that the two models give very similar stellar masses,
and thus that the stellar mass is a robust parameter
against the assumed star formation history.
Dust extinction is also relatively insensitive
to the star formation history.
The only exception is NB570-W-59558,
for which $E(B-V)_\star=0.63$ for constant star formation
and $0.00$ for exponentially decaying star formation.
On the other hand, age and SFR largely differ
between the two models for some objects.
The ages of two objects (NB570-C-24119 and NB570-W-59558)
differ by more than one order of magnitude,
and the SFRs of seven objects (NB503-S-65716,
NB503-S-94275,NB503-N-21105, NB503-$K$-undetected,
NB570-C-24119, NB570-N-32295, NB570-W-59558)
differ by more than one order of magnitude.
All but one of these seven objects have $\tau=0.001$ Gyr,
i.e., burst-like star formation histories.
In other words, these objects are equally well fitted
by two extremes, constant star formation and
burst-like star formation,
and our data cannot discriminate between them.
It will be resolved by deeper NIR imaging 
\citep[e.g.,][]{pozzetti2000} 
and/or other independent observations 
such as submillimetre imaging.

In the following discussion, we will adopt the results
for constant star formation, because
(i) the values of reduced $\chi^2$ are lower for constant star
formation for all but one object and
(ii) it is easy to compare our results
with previous SED studies on LAEs
since most of them have also adopted constant star formation.

Most of the LAEs have a brighter $K$-band
magnitude than the best-fit SED.
The reason for this discrepancy is not clear, but
a possible reason is that the $K$-band flux density is
affected by [O\,{\sevensize III}] and H$\beta$ emission.
On the basis of the relations between SFR and [O\,{\sevensize III}] and
H$\beta$ emission strengths \citep{kennicutt1998,moustakas2006},
we find that these lines may significantly increase $K$-band flux
densities if SFR is higher than $\sim 10^3$ [$M_\odot$ yr$^{-1}$].
However, since most of the objects have SFR $\la 10^3 M_\odot$ yr$^{-1}$,
this effect cannot explain the observed excesses.
Another possible reason is that
a relatively faint magnitude, $K(3\sigma)=24.1$, is adopted
for the boundary of $K$-detection.
Our $K$-detected sample may include objects whose $K$ magnitude
happens to be brighter than the true value due to positive
sky noise at the position of the objects.
However, most of the $K$-detected objects are also detected in at
least one of the $J$, $H$, and the four IRAC bands,
suggesting that they are truly bright in the near-infrared wavelengths
and thus that the systematic brightening of $K$ is not very strong.
We infer that the systematic brightening of $K$, if any,
does not significantly affect the best-fit SEDs which are determined
from the overall shapes of observed SEDs
from $R$ to IRAC $8.0 \mu$m bands.
Indeed, we perform the SED fitting excluding $K$ photometry,
and find that for any object the best-fit parameters which are
different from the original values only within the $1\sigma$
errors, although the best-fit stellar masses tend
to be slightly ($\simeq 0.05$ dex) smaller.

Most of the errors in the best-fit parameters of NB503-$K$-undetected
are larger than those of NB570-$K$-undetected
in spite of a larger number of objects being stacked in NB503.
This is because NB503-$K$-undetected is fainter in most bandpasses
as shown in Table 5.
The $z=3.1$ LAE sample includes fainter objects than the $z=3.7$
sample since the limiting magnitude in NB503 is fainter
by $0.6$ mag \cite[see Section 2.2 in][]{ouchi2008}.
Thus, the $K$-undetected objects in the $z=3.1$ sample are on average
fainter than those in the $z=3.7$ sample.
We also note that the reduced $\chi^2$ is smaller for
NB503-$K$-undetected.
We infer that this is 
not only because the observed $K$-band flux density for 
NB570-$K$-undetected is much brighter than the best-fit SED 
(see Figure \ref{fig:SEDs2}), 
but also because the magnitude errors of NB503-$K$-undetected 
are larger.

Some new population synthesis models include
thermally pulsating asymptotic giant branch (TP-AGB) stars
\citep[e.g.,][]{maraston2005,bruzual2007}.
Inclusion of TP-AGB stars should have little effect on
the SED fitting for most LAEs,
because LAEs are in general very young and have subsolar metallicities
\citep[e.g.,][]{pentericci2008,gawiser2009}.
In our sample, however, four objects
(NB503-N-42377, NB503-S-45244, NB570-W-55371, and NB570-C-24119)
have relatively old ages of $> 300$ Myr and thus their NIR fluxes
could be dominated by TP-AGB stars.
We fit a new version of GALAXEV which includes TP-AGB stars
(Charlot {\&} Bruzual, in preparation) to these four,
and obtain almost the same results as for BC03,
except decreases in stellar mass of $0.2$ -- $0.3$ dex.


\section{Results and Discussion} \label{sec:results}

\begin{figure*}
 \begin{center}
  \includegraphics[scale=0.9]{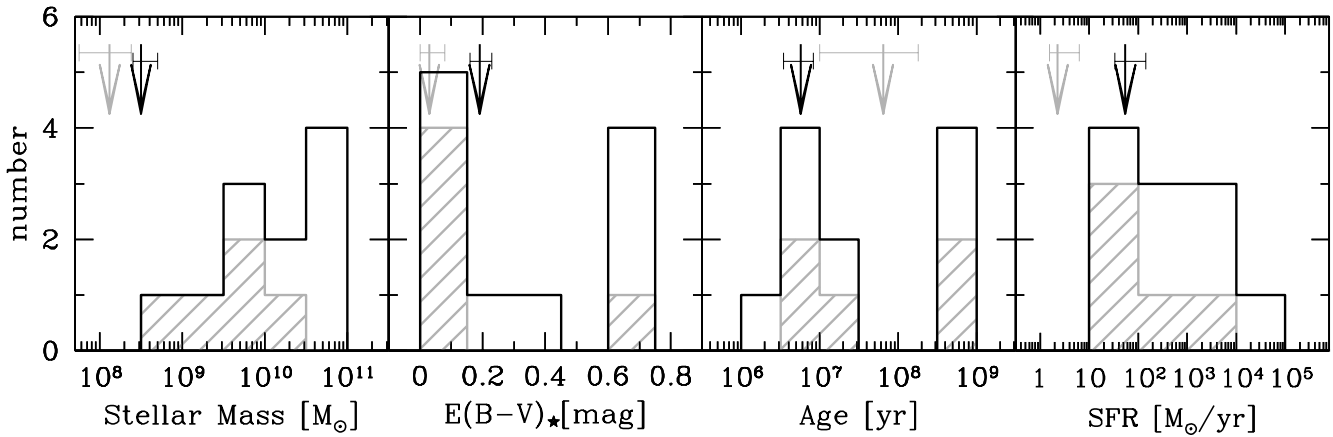}
 \end{center}
 \caption[
The distributions of the best-fit parameters
derived from our SED fitting.
]{
Distributions of the best-fit parameters
derived from the SED fitting.
The shaded histograms are for the $K$-detected LAEs at $z=3.1$,
and the open histograms are  for the sum of the $K$-detected LAEs at $z=3.1$ and $3.7$.
Grey and black arrows indicate
the $K$-undetected LAEs at $z=3.1$ and $3.7$, respectively.
}
\label{fig:histograms}
\end{figure*}

Figure \ref{fig:histograms} shows
the distributions of the best-fit parameters
derived from the SED fitting.
The stellar masses of the $K$-detected LAEs
are distributed over
$M_{\rm star} \sim 10^9 $ -- $10^{10.5} M_\odot$,
while those of the $K$-undetected LAEs
are much less massive
($M_{\rm star} \sim 10^8 $ -- $10^{8.5} M_\odot$).
The $K$-detected LAEs have
a large variety of dust extinction
with $0 < E(B-V)_\star < 0.7$,
while the $K$-undetected LAEs have mild dust extinction.
The ages of our LAEs are distributed
over around $10^6$ -- $10^9$ yr.
The age distribution of the $K$-detected objects could be
bimodal, although its statistical significance is low.
The ages of the $K$-undetected LAEs
are around the median of the $K$-detected LAEs.
The SFRs of the $K$-detected LAEs
span a wide range of $10$ -- $10^4 M_\odot$ yr$^{-1}$,
while those of the $K$-undetected LAEs
are between $1$ -- $100 M_\odot$ yr$^{-1}$.

The $E(B-V)_\star$ of NB570-$K$-undetected is significantly
larger than that of NB503-$K$-undetected.
This may merely reflect the fact that NB570-$K$-undetected is
more massive and thus has produced more metal.
On the other hand, this could reflect some evolution of the LAE
population between the two redshifts.
Indeed, it is found in Table \ref{tab:sum_stellarpop} that
LAEs at $z>3.7$ can have $E(B-V)_\star$ comparable to 
or larger than that of NB570-$K$-undetected.

We find that four objects have extremely large extinction $E(B-V)_\star \sim 0.65 $:
NB503-N-21105, NB570-N-32295, NB570-S-88963, and 
NB570-W-59558\footnote{This object is equally well fitted by two extremes:  
a young and dusty starburst galaxy with constant SFH, 
and an old and dead galaxy with exponentially decaying SFH. 
This is caused by the age-dust degeneracy.  
Although our data cannot discriminate between them, 
we treat this object as a young and dusty starburst galaxy 
in this section (see details in Section \ref{subsec:SEDfitting-results})}.
Their properties (other than age)
are clearly different from the other LAEs,
as we can see in the following subsections.
We label the four objects as R1 -- R4
in most of the forthcoming figures.
These four red LAEs may be a distinct population,
and we will discuss this in Section \ref{subsec:red-laes}.

\subsection{Correlations among Stellar Population Parameters}

Table \ref{tab:sum_stellarpop} summarises the results of
the stellar population analysis of LAEs reported so far.
In this subsection, we examine correlations
among stellar population parameters,
referring to the previous studies.

\begin{table*}
\centering
\caption{Summary of Stellar Populations}
 \label{tab:sum_stellarpop}
{\scriptsize
\begin{tabular}{ccccccccc}
\hline
Reference
& $z$
& $M_{\rm star}$
& $E(B-V)_\star$
& Age
& $Z$
& SFR
& $\tau$
& Remarks
\\

&
& [$10^9 M_\odot$]
& [mag]
& [Myr]
& [$Z_\odot$]
& [$M_\odot$ yr$^{-1}$]
& [Myr]
&
\\
\hline
This Study
 &  $3.1$
 &  $0.13$
 &  $0.03$
 &  $65$
 &  ($0.2$)
 &  $2.3$
 &  ($\infty$)
 &  (O09-1)
\\
This Study
 &  $3.1$
 &  $0.93$ -- $27$
 &  $0.00$ -- $0.70$
 &  $4.8$ -- $407$
 &  ($0.2$)
 &  $11$ -- $5.6 \times 10^3$
 &  ($\infty$)
 &  (O09-2)
\\
\cite{gawiser2006}
 &  $3.1$
 &  $0.5$
 &  $0.0$
 &  $90$
 &  ($1$)
 &  $6$
 &  ($\infty$)
 &
\\
\cite{gawiser2007}
 &  $3.1$
 &  $1.0$
 &  $0.0$
 &  $20$
 &  ($1$)
 &  $2$
 &  $750$
 &  (G07)
\\
\cite{lai2008}
 &  $3.1$
 &  $0.3$
 &  $0.0$
 &  $1.6 \times 10^2$
 &  ($1$)
 &  $2$
 &  ($\infty$)
 &  (L08-1)
\\
\cite{lai2008}
 &  $3.1$
 &  $9$
 &  $0.0$
 &  $1.6 \times 10^3$
 &  ($1$)
 &  $6$
 &  ($\infty$)
 &  (L08-2)
\\
\cite{nilsson2007}
 &  $3.1$
 &  $0.47$
 &  $0.07$
 &  $8.5 \times 10^2$
 &  ($0.005$)
 &  $0.66$
 &  ($\infty$)
 &
\\
This Study
 &  $3.7$
 &  $0.32$
 &  $0.19$
 &  $5.8$
 &  ($0.2$)
 &  $55$
 &  ($\infty$)
 &  (O09-1)
\\
This Study
 &  $3.7$
 &  $3.9$ -- $51$
 &  $0.04$ -- $0.67$
 &  $1.4$ -- $9.1 \times 10^2$
 &  ($0.2$)
 &  $21$ -- $3.5 \times 10^4$
 &  ($\infty$)
 &  (O09-2)
\\
\cite{finkelstein2008b}
 &  $4.5$
 &  $0.084$ -- $6.1$
 &  $0.02$ -- $0.41$
 &  $2.5$ -- $500.0$
 &  $0.005$ -- $1$
 &  $1.30$ -- $5.78$
 &  $0.1$ -- $4 \times 10^3$
 &  (F09)
\\
\cite{pirzkal2007}
 &  $4.00$ -- $5.76$
 &  $0.007$ -- $1.8$
 &  $0.0$ -- $0.16$
 &  $1.0$ -- $20$
 &  $0.005$ -- $2.5$
 &  ---
 &  $1$ -- $12.5$
 &  (P07)
\\
\cite{lai2007}
 &  $5.7$
 &  $17$ -- $39$
 &  $0.150$ -- $0.225$
 &  $720$ -- $900$
 &  ($0.005$)
 &  ---
 &  $(\infty)$
 &  (L07)
\\
\cite{chary2005}
 &  $6.56$
 &  $0.84$
 &  $0.25$
 &  $5$
 &  $0.02$
 &  $140$
 &  ---
 &  (C05)
\\
\hline
\end{tabular}
} 

\medskip
\begin{minipage}{170mm}
\begin{flushleft}
\textbf{NOTES}:
Values in parentheses are the assumed values in the literature.

\textbf{REMARKS}:
\textbf{(O09-1)} $K$-undetected LAEs.
\textbf{(O09-2)} $K$-detected LAEs.
\textbf{(G07)} Their model has two stellar population components
with different ages, and we adopt here the young component.
The age of the old component is $2$ Gyr.
\textbf{(L08-1)} IRAC-undetected LAEs.
\textbf{(L08-2)} IRAC-detected LAEs.
\textbf{(P07)} The result with an exponentially decaying star formation history.
In Figure \ref{fig:Mstar_sSFR}, we show the results shown in Figure 4 of \cite{castro2008}.
\textbf{(F09)} The result with an exponentially decaying star formation history.
SFRs are calculated from model SEDs.
\textbf{(L07)} In Figure \ref{fig:Mstar_sSFR},
we show their results based on the assumption of constant star formation history and $0.005 Z_\odot$.
\textbf{(C05)} They calculate the SFR from the H$\alpha$ luminosity.
\end{flushleft}
\end{minipage}
\end{table*}

\subsubsection{$M_{\rm 1500} - M_{\rm V}$ versus $M_{\rm 1500}$ }

\begin{figure}
 \begin{center}
  \includegraphics[scale=0.45]{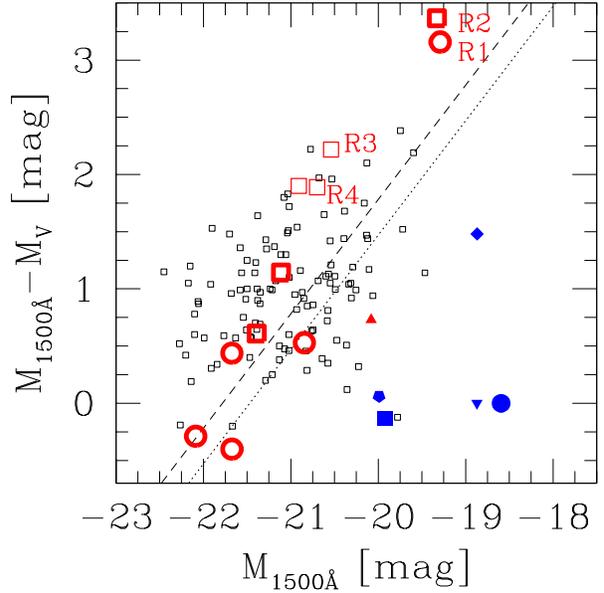}
 \end{center}
 \caption[UV-to-optical colour ($M_{\rm 1500} - M_{\rm V}$)
versus
rest-frame UV absolute magnitude ($M_{\rm 1500}$) ]
{
UV-to-optical colour ($M_{\rm 1500} - M_{\rm V}$)
versus 
rest-frame UV absolute magnitude ($M_{\rm 1500}$)
for LAEs at $z=3.1$ and $3.7$,
and LBGs at $z \sim 3$.
The red circles and squares indicate the $K$-detected LAEs
at $z = 3.1$ and $3.7$, respectively,
where those with spectroscopic confirmation
are shown by bold symbols.
Labels R1 -- R4 indicate
$K$-detected LAEs with extremely large
dust extinction ($E(B-V)_\star \sim 0.65$).
The blue filled circles and squares represent the $K$-undetected LAEs
at $z = 3.1$ and $3.7$, respectively.
The dotted and dashed lines correspond to
the $3\sigma$ detection limit in the $K$ band
for LAEs at $z=3.1$ and $z=3.7$, respectively.
The black squares are LBGs at $z \sim 3$
\citep{shapley2001,papovich2001,iwata2005},
all of which are $K$-detected and
spectroscopically confirmed.
The blue upside-down pentagon and diamond represent
composite LAEs at $z\sim3.1$ \citep{gawiser2006}
and $z\sim3.15$ \citep{nilsson2007}.
The red and blue triangles indicate
stacked IRAC-detected and undetected LAEs at $z=3.1$ \citep{lai2008}.
}
\label{fig:R_RK_z3_all}
\end{figure}

Figure \ref{fig:R_RK_z3_all} plots in large symbols
rest-frame UV-to-optical colour ($M_{1500} - M_{\rm V}$)
against UV($= 1500${\AA}) absolute magnitude 
($M_{1500}$) for our LAEs.
For conservative discussion, rest-frame
$1500${\AA} and
$V$-band absolute magnitudes 
of our sample are 
derived from the best-fit SEDs,
not from observed
$R$- and $K$-band photometry
(see details in Section \ref{subsec:SEDfitting-results}).

For the $11$ $K$-detected LAEs, nine are distributed
in a similar region in the $M_{1500} - M_{\rm V}$
versus $M_{1500}$ plane to the LBGs.
The remaining two are faint in $M_{1500}$ and
extremely red ($M_{1500} - M_{\rm V} > 3$).
Indeed, these are two of the four dusty LAEs, R1 and R2.
On the contrary, the $K$-undetected LAEs, which occupy $96$ {\%}
of our sample, have bluer $M_{1500}-M_{\rm V}$ colours and much
fainter $M_{1500}$ magnitudes than the LBGs.
The existence of $K$-bright LAEs suggests that not all LAEs
at $z\sim 3$ are low-mass galaxies with blue colours
and that
LAEs are a heterogeneous population with wide ranges
of stellar mass, age, and/or dust extinction
\cite[e.g.,][]{lai2008,finkelstein2008a,finkelstein2008b,pentericci2008}.

We also compare our LAEs
with those in the literature at similar redshifts.
The IRAC-detected LAEs by \cite{lai2008}
have similar $M_{1500} - M_{\rm V}$ and $M_{1500}$ to the LBGs.
On the other hand, the NIR-undetected LAEs
by \cite{gawiser2007} and \cite{lai2008}
have as blue $M_{1500}-M_{\rm V}$ colours
and faint $M_{1500}$ magnitudes
as our $K$-undetected LAEs.
The NIR-undetected LAE by \cite{nilsson2007}
has redder $M_{1500} - M_{\rm V}$ than
the other NIR-undetected LAEs,
probably because their $K$-band image is relatively shallow
(Table \ref{tab:summary_of_earliers}).

\subsubsection{Stellar Mass versus $M_{\rm V}$ and $M_{\rm 1500}$}

\begin{figure*}
  \includegraphics[scale=0.9]{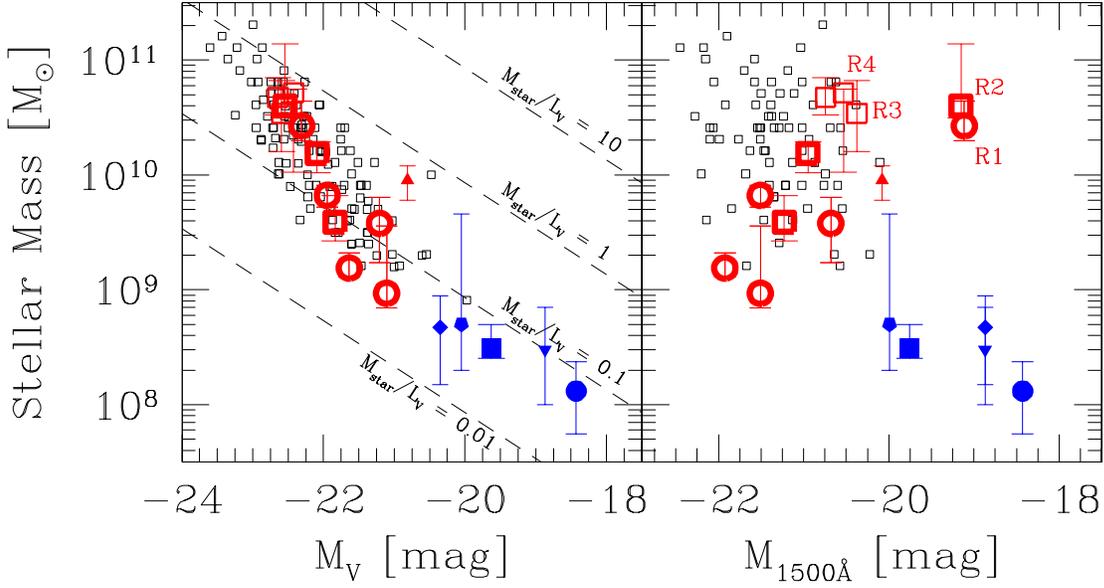}
 \caption{
\textbf{Left}:
Stellar mass versus
rest-frame $V$-band absolute magnitude.
All symbols are the same as in Figure \ref{fig:R_RK_z3_all}.
Black dashed lines correspond to
four mass-to-luminosity ratios
normalised by the solar value. 
\textbf{Right}:
Stellar mass versus rest-frame UV absolute magnitude.
All symbols are the same as in Figure \ref{fig:R_RK_z3_all}.
}
\label{fig:restOptUV_Mstar}
\end{figure*}

The left panel of Figure \ref{fig:restOptUV_Mstar} shows stellar mass
as a function of rest-frame $V$-band absolute magnitude
for our LAEs calculated from best-fit SEDs.
As expected, the $K$-detected LAEs have comparable
stellar masses to the $K$-detected LBGs, spanning
over $M_{\rm star} \sim 10^9$ -- $10^{10.5} M_\odot$,
while the $K$-undetected LAEs are much less massive,
with $10^8$ -- $10^{8.5} M_\odot$.
Our results broadly agree with previous studies.
This figure shows that LAEs have a wide range of stellar mass.

The dashed lines indicate four mass-to-luminosity ratios.
It is found that our $K$-undetected LAEs have slightly
lower $M_{\rm star}/L_{\rm V}$ ratios than
the average $K$-detected LAEs and LBGs,
suggestive of younger ages.

The right panel of Figure \ref{fig:restOptUV_Mstar}
plots stellar mass against rest-frame UV absolute magnitude.
All but two of the $K$-detected LAEs are distributed
in a region similar to the LBGs.
The two exceptions are R1 and R2.
The very faint $M_{\rm 1500}$ magnitudes
with respect to high $M_{\rm star}$ of
R1 and R2 are mainly due to
the extremely large dust extinction
(see Table \ref{tab:SEDfitting}).
The $K$-undetected LAEs are much fainter in UV wavelengths
and much less massive than the LBGs.

\subsubsection{Star Formation Rate versus $M_{\rm 1500}$ and $L({\rm Ly}\alpha)$}

\begin{figure*}
 \begin{center}
  \includegraphics[scale=0.9]{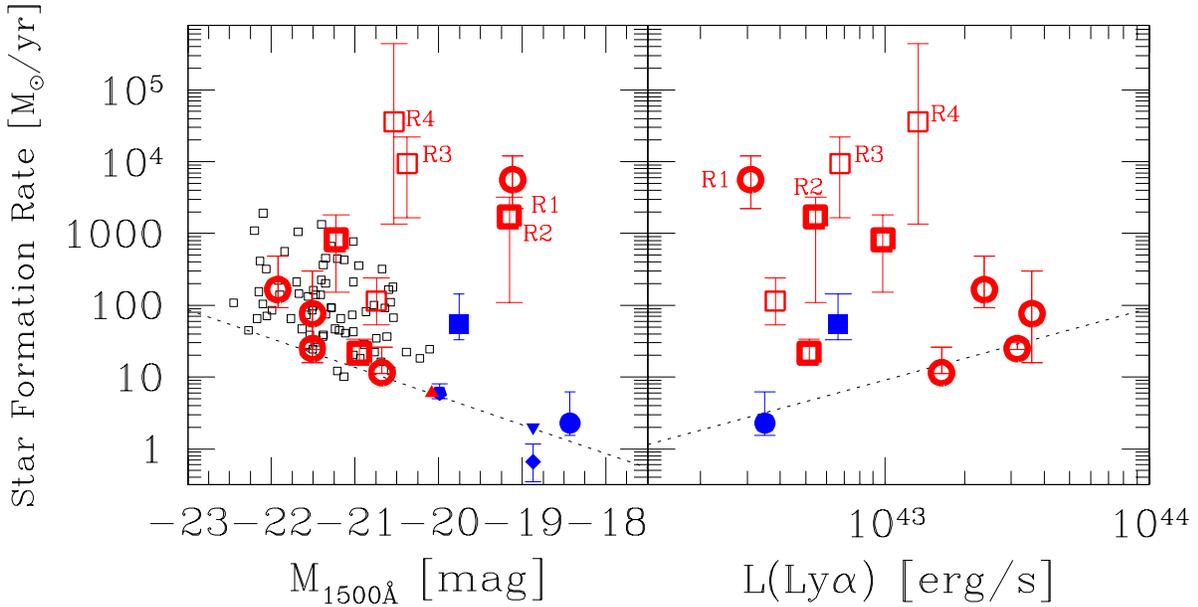}
 \end{center}
 \caption[
Star formation rate derived from SED fitting
versus
rest-frame UV absolute magnitude.
]
{
\textbf{Left}:
Star formation rate derived from SED fitting, plotted against
the UV absolute magnitude.
All symbols are the same as in Figure \ref{fig:R_RK_z3_all}.
The dotted line shows the relation between the UV luminosity and
the star formation rate \citep{madau1998}.
\textbf{Right}:
Star formation rate derived from SED fitting, plotted against
the Ly$\alpha$ luminosity.
All symbols are the same as in Figure \ref{fig:R_RK_z3_all}.
The dotted line shows the relation between the Lyman $\alpha$ luminosity and
the star formation rate.
}
\label{fig:restUVLLyA_SFR}
\end{figure*}

The left panel of Figure \ref{fig:restUVLLyA_SFR}
plots the star formation rate against the rest-frame UV absolute magnitude.
For all objects SFRs are calculated from the best-fit SEDs.
For the case of $E(B-V)_\star = 0$, SFR approximately correlates with $L_{\rm 1500}$,
and we plot this correlation with the dotted line using \citep{madau1998}:
\begin{equation}
{\rm SFR} \, [M_\odot {\rm yr}^{-1}]
 = \frac{ L_{\rm 1500} [{\rm erg}\,\,{\rm s}^{-1} {\rm Hz}^{-1}] }{8 \times 10^{27}}.
\label{eq:madau1998}
\end{equation}

Reflecting the wide range of $E(B-V)_\star$,
the $K$-detected objects are widely scattered above the dotted line.
Some objects with very low extinction are
located on the dotted line,
while the four objects with extremely large extinction
(R1 -- R4) have SFRs more than two orders of magnitude
higher than those expected from the observed $M_{\rm 1500}$.
Except for R1 -- R4, the $K$-detected LAEs
have similar ranges of SFR and $M_{\rm 1500}$ to the LBGs.

The $K$-undetected LAEs have SFRs of $1$ -- $100$ $M_\odot$ yr$^{-1}$
and their offsets from the dotted line are within an order of magnitude
because of the modest dust extinction.
NB570-$K$-undetected is offset nearly an order of magnitude
because of its relatively heavy dust extinction (Table \ref{tab:SEDfitting}).

The right panel of Figure \ref{fig:restUVLLyA_SFR} shows
star formation rate versus observed Lyman $\alpha$ luminosity.
Based on the relation between H$\alpha$ luminosity
and star formation rate \citep{kennicutt1998}
under case B approximation \citep{brocklehurst1971},
star formation rate is related with
Lyman $\alpha$ luminosity as:
\begin{equation}
{\rm SFR}\, [M_\odot {\rm yr}^{-1}]
 = 9.1 \times 10^{-43} L ({\rm Ly}\alpha) [{\rm erg}\,\,{\rm s}^{-1}] .
\label{eq:kennicutt1998}
\end{equation}
The dotted line shows this relation.
Star formation rates derived from SED fitting
are found to be higher than those from equation (\ref{eq:kennicutt1998}).
This can be explained by several reasons.
Ly$\alpha$ photons can be attenuated by dust,
and scattered in the interstellar and intergalactic medium.
Ly$\alpha$ photons are resonantly scattered and heavily attenuated
especially under a homogeneous ISM.
Intrinsically, the Ly$\alpha$ emissivity is also affected by
the age, metallicity, and IMF.
In addition, the $L ({\rm Ly}\alpha)$ of objects
without spectroscopic redshifts tends to be
underestimated due to the triangle shapes of the narrow-band filters.
NB570-$K$-undetected lies well above equation (\ref{eq:kennicutt1998}), 
probably due to its relatively large $E(B-V)_\star$.

\subsubsection{Stellar Mass, Age, and $E(B-V)_\star$ versus EW(Ly$\alpha$)}

\begin{figure}
 \begin{center}
  \includegraphics[scale=0.94]{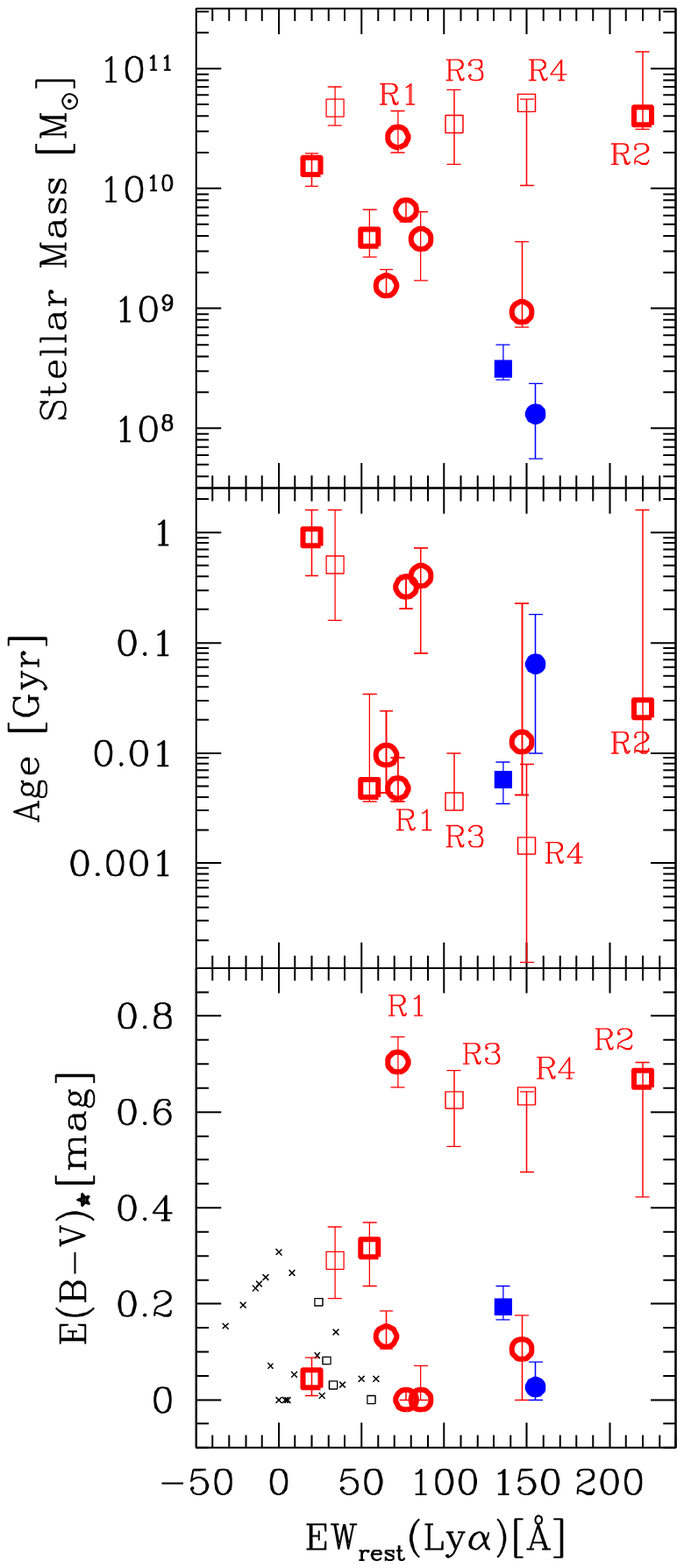}
 \end{center}
 \caption[]{
\textbf{Top}:
Stellar mass derived from SED fitting versus
rest-frame Ly$\alpha$ equivalent width. 
All symbols are the same as in Figure \ref{fig:R_RK_z3_all}.
\textbf{Middle}:
Stellar age derived from SED fitting
versus rest-frame Ly$\alpha$ equivalent width. 
All symbols are the same as in Figure \ref{fig:R_RK_z3_all}.
\textbf{Bottom}:
Dust reddening $E(B-V)_\star$ derived from SED fitting
versus rest-frame Ly$\alpha$ equivalent width.
The black small squares are taken from \cite{pentericci2008}.
The crosses are star-forming galaxies in the local universe 
taken from \cite{giavalisco1996}.
The remaining symbols are the same as in Figure \ref{fig:R_RK_z3_all}.
}
\label{fig:EW_Pops}
\end{figure}

The top panel of Figure \ref{fig:EW_Pops} shows
stellar mass versus EW(Ly$\alpha$).
The $K$-detected objects are massive and
have a wide range of EW(Ly$\alpha$) over $20$ -- $220$ {\AA},
but mostly in the range $\la 100$ {\AA}.
On the other hand,
the $K$-undetected objects are less massive and
have consistently high values of $\sim 150$ {\AA}.
Considering that the $K$-undetected objects closely represent
the whole LAE population, we can conclude from this figure that
massive LAEs with high EW(Ly$\alpha$) are rare.

There is a weak anti-correlation between stellar mass and EW(Ly$\alpha$),
if the four red objects (R1 -- R4) are excluded.
This anti-correlation is qualitatively consistent with the
finding by \cite{pentericci2008} in their LAE sample
from $z \sim 3.5$ to $6$
that more massive objects have in general
smaller EW(Ly$\alpha$).

This anti-correlation is not due to the Lyman $\alpha$ emission of
more massive objects being more heavily absorbed by dust.
For example, NB570-W-55371 has a small extinction ($E(B-V)_\star = 0.04$)
and a small EW(Ly$\alpha$) ($20$ {\AA}),
while NB503-S-94275 with a moderate extinction ($E(B-V)_\star = 0.11$)
has a large EW(Ly$\alpha$) of $150$ {\AA}
and the $z=3.7$ $K$-undetected objects with $E(B-V)_\star = 0.19$
have a large EW(Ly$\alpha$) of $130$ {\AA}.
Indeed, the bottom panel of Figure \ref{fig:EW_Pops}
shows no significant correlation
between $E(B-V)_\star$ and EW(Ly$\alpha$).
A cause of the anti-correlation may be that the geometry of dust
in LAEs changes with stellar mass in the sense that more
massive galaxies have a less clumpy distribution for some reason.
As \cite{neufeld1991} points out,
the fraction of Ly$\alpha$ photons
escaping from a galaxy is higher for a clumpy cloud distribution
than for a uniform distribution.
This effect can explain the observed SEDs of LAEs at $z \sim 4.5$
\citep{finkelstein2008b}.

The middle panel of Figure \ref{fig:EW_Pops} shows
stellar age versus EW(Ly$\alpha$).
There seems to be a weak anti-correlation between
stellar age versus EW(Ly$\alpha$).

In the bottom panel of Figure \ref{fig:EW_Pops},
the small squares indicate the
four averaged values
of $38$ LAEs at $z \sim 3.5$ -- $6$ by \cite{pentericci2008}.
A probable reason for their very small $E(B-V)_\star$ is
that they have been selected using the LBG technique
and thus have blue colours.
The crosses represent the data for local star forming galaxies
given in \cite{giavalisco1996}.
In comparison to our LAEs,
these local galaxies have on average smaller EWs.
This might suggest a more uniform cloud distribution for local galaxies,
if \cite{neufeld1991} picture is correct.

\subsubsection{Ly$\alpha$ escape fraction versus $E(B-V)_{\rm gas}$}

\begin{figure}
 \begin{center}
  \includegraphics[scale=0.4]{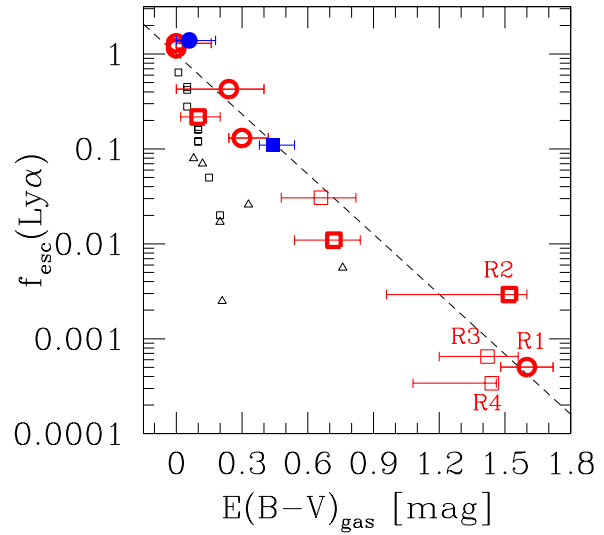}
 \end{center}
 \caption[
Ly$\alpha$ escape fraction versus $E(B-V)$.
]
{
Ly$\alpha$ escape fraction versus $E(B-V)_{\rm gas}$.
As for our results, $E(B-V)_{\rm gas} = E(B-V)_\star / 0.44$ \citep{calzetti2000}.
The triangles are star-forming galaxies
in the local universe \citep{atek2008},
and the black small squares are $z \sim 3$ LBGs \citep{verhamme2008}.
The remaining symbols are the same as in Figure \ref{fig:R_RK_z3_all}.
The dashed line corresponds to equation (\ref{eq:fesc_EW_EBV}).
}
\label{fig:EBV_fesc}
\end{figure}

We define the Ly$\alpha$ escape fraction
$f_{\rm esc} ({\rm Ly}\alpha)$ by:
\begin{equation}
f_{\rm esc}({\rm Ly}\alpha)
      = \frac{L_{\rm obs}({\rm Ly}\alpha) [{\rm erg}\,\,{\rm s}^{-1}]}{L_{\rm int}({\rm Ly}\alpha) [{\rm erg}\,\,{\rm s}^{-1}]},
\label{eq:f_esc}
\end{equation}
where $L_{\rm int}({\rm Ly}\alpha)$ [erg s$^{-1}$]
$= 1.1 \times 10^{42}\,{\rm SFR}\,[M_\odot {\rm yr}^{-1}]$,
equivalent to equation (\ref{eq:kennicutt1998}),
is the intrinsic Ly$\alpha$ luminosity
computed from the SFR on the assumption of case B.
Since the SFR is derived from our SED fitting, it has been
corrected for dust extinction.
Then $f_{\rm esc} ({\rm Ly}\alpha)$ is plotted against $E(B-V)_{\rm gas}$
in Figure \ref{fig:EBV_fesc}.
As for our results, $E(B-V)_{\rm gas} = E(B-V)_\star / 0.44$ \citep{calzetti2000}.
A strong negative correlation is seen between the two quantities
for our objects.
The $K$-undetected LAEs, whose dust extinction is modest, have
relatively large $f_{\rm esc} ({\rm Ly}\alpha)$ of
$\simeq 0.1$ -- $1$, while the $K$-detected LAEs, whose
$E(B-V)_{\rm gas}$ can be as large as $1.5$, have
$f_{\rm esc} ({\rm Ly}\alpha) \sim 3 \times 10^{-4}$ -- $1$.
R1 -- R4 have the smallest escape fractions.

We discuss implications of this negative correlation.
Assuming $f_\nu =$ constant over $1200$ -- $1500$ {\AA},
we obtain
$L_{\rm obs}({\rm Ly}\alpha) \equiv
{\rm EW}({\rm Ly}\alpha) (c/\lambda_0^2) L_{1500,{\rm obs}}$ as:
\begin{equation}
L_{\rm obs}({\rm Ly}\alpha)
 = {\rm EW}({\rm Ly}\alpha)
  \frac{c}{\lambda_0^2} L_{1500} 10^{-0.4R(\lambda_0)E(B-V)_{\rm gas}},
\end{equation}
where $\lambda_0=1216$ {\AA}, and $R(\lambda_0) = 0.44 \, k'(\lambda_0) \simeq 5.27$ is
the selective extinction at $1216$ {\AA} for Calzetti extinction law.
Combining this equation with equations (\ref{eq:madau1998}) and (\ref{eq:kennicutt1998}),
we rewrite equation (\ref{eq:f_esc}) as:
\begin{equation}
f_{\rm esc}({\rm Ly}\alpha)
      = \frac{8 \times 10^{27}}{1.1 \times 10^{42}}
              \cdot \frac{c}{\lambda_{\rm 0}^2}
              \cdot {\rm EW} ({\rm Ly}\alpha) \cdot 10^{-0.4 R(\lambda_{\rm 0}) E(B-V)_{\rm gas}}
\label{eq:fesc_EW_EBV}
\end{equation}
The dashed line in Figure \ref{fig:EBV_fesc}
corresponds to this equation for EW(Ly$\alpha$) $= 68${\AA},
the value of EW(Ly$\alpha$) expected for $f_{\rm esc}({\rm Ly}\alpha) = 1$
when $E(B-V)_{\rm gas} = 0$.
In other words, we assume here that the absorption fractions
of Ly$\alpha$ photons and continuum photons near Ly$\alpha$
wavelength are identical
irrespective of $E(B-V)_{\rm gas}$.
Clearly, this assumption is not valid for uniform ISM,
since Ly$\alpha$ photons are more heavily absorbed
before escaping from the galaxy.
In this sense, it is interesting that almost all of
our LAEs are on the dashed line.
Similar to the bottom panel of Figure \ref{fig:EW_Pops},
which shows no significant correlation between $E(B-V)_\star$ and EW(Ly$\alpha$),
this results also suggests that in heavily obscured LAEs
the absorption of Ly$\alpha$ photons is not as strong as
expected for uniform ISM.

Our result is consistent with the result of \cite{valls-gabaud1993},
who has studied nearby starburst galaxies
and showed that most of them present normal Ly$\alpha$ emission
consistent with case B recombination.
This implies that the effect of resonant scattering is not so strong.

\cite{finkelstein2008b} have studied  $z\sim4.5$ LAEs considering their dust geometries
by introducing the $q$ parameter as the ratio of the optical depth
for Ly$\alpha$ to that for UV continuum.
Our results suggest that
Ly$\alpha$ photons and UV continuum photons near Ly$\alpha$ wavelength
are equally attenuated by dust.
This corresponds to $q=1$.
Interestingly, \cite{finkelstein2008b} have found that,
although three objects have $q > 1$ and two have $q=1$,
the remaining nine have $q<1$.
This might mean that the dust effect on Ly$\alpha$ photons 
is more important at higher redshift.

The black small squares represent the results for
11 LBGs at $z \sim 3$ by \cite{verhamme2008}.
Although they also show a negative correlation,
its slope is steeper than the dashed line.
This means that for a given $E(B-V)_{\rm gas}$, Ly$\alpha$ photons
are more strongly absorbed in comparison with LAEs.
Similarly, local star-forming galaxies
\citep[open triangles,][]{atek2008} also tend to have lower
$f_{\rm esc}({\rm Ly}\alpha)$.

\subsubsection{Specific Star Formation Rate versus Stellar Mass:
Comparison with other high-$z$ galaxy populations }

\begin{figure*}
 \begin{center}
  \includegraphics[scale=0.7]{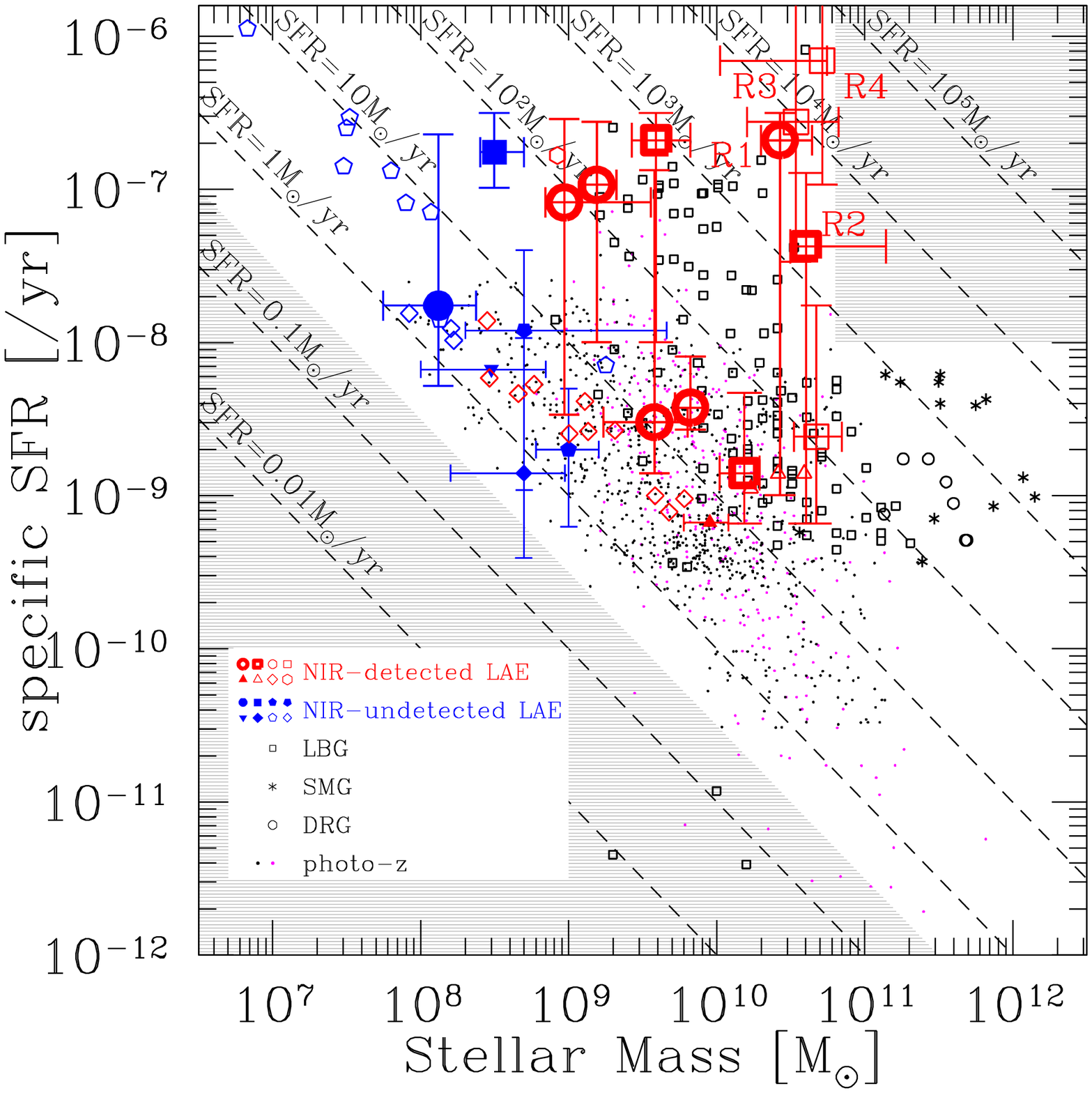}
 \end{center}
 \caption[
Stellar mass versus
specific star formation rate
($=$SFR/stellar mass).
]
{
Stellar mass versus
specific star formation rate
($=$SFR/stellar mass).
The dashed lines represent SFR $= 0.01$ --
$10^5 M_\odot$ yr$^{-1}$ from left to right .
The blue filled
pentagon represents a stacked LAE
of $52$ objects at $z=3.1$ \citep{gawiser2007}.
The red and blue open diamonds are
$14$ LAEs $z \sim 4.5$ with and without IRAC-detection
\cite[private communication]{finkelstein2008b}.
The blue open pentagons are LAEs at $z \sim 5$
\citep{pirzkal2007,castro2008}.
The red open triangles are three IRAC-detected LAEs
at $z=5.7$ \citep{lai2007}.
The red open hexagon is a lensed LAE at $z=6.56$ \citep{chary2005}.
The star marks are SMGs from \cite{borys2005} and \cite{chapman2005}.
The small circles are DRGs from \cite{vandokkum2004}.
The black and magenta dots are 
$I$-band and $K$-band detected photo-$z$ samples respectively \citep{feulner2005}.
The remaining symbols are the same as in Figure \ref{fig:R_RK_z3_all}.
There is no object in the region of the high stellar mass
and sSFR (shaded region in the top-right corner).
The lower-left shaded region 
corresponds to quiescent low-mass objects,
almost all of which are missed so far.
}
\label{fig:Mstar_sSFR}
\end{figure*}

We plot in Figure \ref{fig:Mstar_sSFR} the specific star formation
rate (sSFR $\equiv$ SFR/$M_{\rm star}$) against the stellar mass
for our LAEs and other high-redshift galaxies
selected by various photometric selection methods,
in order to compare the mass scale and
star-formation activity between the high-$z$ galaxy populations.
Also plotted are LAEs over a wide range of redshift taken from the literature.
The data plotted are summarised below:

\vspace{5pt}
\noindent
\underline{LAEs}

\noindent
Large blue filled circle and large blue filled square:
$K$-undetected (stacked) LAEs in our sample at $z=3.1$ and $3.7$,
respectively.
Large red open circles and large red open squares:
$K$-detected LAEs in our sample at $z=3.1$ and $3.7$, respectively,
where those with a spectroscopic redshift are shown by bold symbols.
Blue filled inverse pentagon: $18$ stacked LAEs at $z=3.1$ \citep{gawiser2006}.
Blue filled diamond: $23$ stacked LAEs at $z = 3.15$ \citep{nilsson2007}.
Blue filled pentagon: $52$ stacked LAEs at $z=3.1$ \citep{gawiser2007}.
Blue filled triangle and red filled triangle:
$76$ IRAC-undetected and $18$ IRAC-detected LAEs at $z=3.1$, respectively \citep{lai2008}.
Red open triangles: three IRAC-detected LAEs at $z=5.7$ \citep{lai2007}.
Blue open diamonds and red open diamonds:
IRAC-undetected and IRAC-detected LAEs at $z=4.5$,
respectively \citep{finkelstein2008b}.
Blue open pentagons: nine LAEs at $z \sim 5$ \citep{pirzkal2007}.
Red open hexagon: a LAE at $z=6.56$ \citep{chary2005}.

\vspace{5pt}
\noindent
\underline{LBGs}

\noindent
Small open squares: $K$-detected LBGs
with spectroscopic redshifts at $z \sim 3$
(Shapley et al. 2001: $K_s < 24.3$,
Papovich et al. 2001: $K_s <23.8$,
Iwata et al. 2005: $K<25.4$).
The distribution of their stellar masses has a lower cutoff
at $\sim 10^9 M_\odot$,
which roughly corresponds to the limiting $K$ magnitudes.

\vspace{5pt}
\noindent
\underline{DRGs}

\noindent
Small open circles: DRGs
at $z \sim 2$ -- $3$ selected by $J-K>1.3$
\citep{vandokkum2004},
which are thought to be either
passively evolving massive galaxies
or dusty starburst galaxies.
Since they are all $K$-bright, their stellar masses
are higher than $\sim 10^{11} M_\odot$.

\vspace{5pt}
\noindent
\underline{SMGs}

\noindent
Asterisks: submillimetre galaxies (SMGs)
at $z \sim 2$ \citep{borys2005,chapman2005},
which are thought to be massive, dusty starburst galaxies.

\vspace{5pt}
\noindent
\underline{Galaxies with Photometric Redshifts}

\noindent
Black and magenta dots: $I$-band and $K$-band selected galaxies
at $2.5 < z < 3.5$, respectively, whose redshifts are estimated
from the photometric redshift technique
\citep{feulner2005}\footnote{\texttt{http://www.pik-potsdam.de/members/feulner/}\\\texttt{research/cosmology/electronic-data}} .

\vspace{5pt}

We first focus on the stacked LAEs (blue filled symbols), which
should closely represent the whole LAE population.
The stellar masses of the stacked LAEs span over
$10^8$ -- $10^9 M_\odot$
(the individually measured LAEs in \cite{pirzkal2007}
have slightly lower masses),
while those of the LBGs,
DRGs,
SMGs,
and photo-$z$ galaxies are
$10^9$ -- $10^{11} M_\odot$,
$10^{11}$ -- $10^{12} M_\odot$,
$10^{11}$ -- $10^{12} M_\odot$,
and $10^{8.5}$ -- $10^{11} M_\odot$, respectively.
Thus,
LAEs are the least massive galaxy population among those discussed here.
This implies that a significant fraction of low-mass galaxies
beyond the limits of broad-band surveys have Lyman $\alpha$
emission strong enough to be detected in narrow-band surveys.
Note, however, that this does not rule out the existence of
low-mass galaxies with weak or without Lyman $\alpha$ emission.

The sSFRs of the stacked LAEs span over two orders of magnitude,
from $10^{-9}$ to $10^{-7}$ yr$^{-1}$, although each measurement
has large uncertainties.
This range of sSFR is similar to that of the LBGs,
implying that these two galaxy populations have similar growth
rates of stellar mass in spite of the large difference
in $M_{\rm star}$.
In contrast, the stacked LAEs appear to have higher sSFRs than
the DRGs, SMGs, and photo-$z$ selected galaxies.
Since the shaded region has not been explored,
the existence of low-mass galaxies with very low sSFRs
is not ruled out.

We then discuss the $K$- and IRAC-detected LAEs.
Our $K$-detected LAEs are distributed in a similar region
to the LBGs (all are $K$-detected), except for the four massive
LAEs with very high sSFRs labelled as R1 -- R4.
\cite{finkelstein2008b}'s IRAC-detected LAEs
are distributed in a similar region
to the photo-$z$ galaxies,
with relatively low stellar masses,
which reflects the deep IRAC Ch1 limit.
On the other hand, \cite{lai2007}
IRAC-detected LAEs at $z=5.7$
are as massive as the LBGs due to
the relatively shallow limiting magnitude.
Overall, the $K$- and IRAC-detected LAEs are not discriminated
from the LBGs and photo-$z$ galaxies
in the sSFR versus $M_{\rm star}$ plane.

\subsection{Properties of Red LAEs}\label{subsec:red-laes}

As seen in the previous subsection, 
the four red LAEs (R1 -- R4) behave quite differently
from the other LAEs in most of the two-parameter spaces.
They have large stellar masses and high SFRs
for relatively faint UV absolute magnitudes
(Figure \ref{fig:restUVLLyA_SFR}).
They deviate from a negative correlation
between $M_{\rm star}$ and EW(Ly$\alpha$)
seen in the majority of LAEs
(top panel of Figure \ref{fig:EW_Pops}).
They have large EWs in spite of extremely large $E(B-V)_\star$
(bottom panel of Figure \ref{fig:EW_Pops}).
Finally, they have much higher sSFRs than
other galaxies with similar stellar masses
(Figure \ref{fig:Mstar_sSFR}).
These results collectively suggest that they are not
just massive LAEs, but may belong to a special class
of galaxies involving obscured starburst.
In this subsection, we further investigate their properties
in conjunction with MIPS $24 \mu$m data and the IMACS spectra.

\subsubsection{Star Formation Rates based on MIPS Photometry}

All four objects have been imaged with MIPS
at $24 \mu$m in the SpUDS.
We find that R4 (NB570-W-59558) is detected
at $\simeq 5\sigma$ level ($m_{24\mu {\rm m}} = 19.3$)
while the remaining three are fainter
than the $3\sigma$ limit ($m_{24\mu {\rm m}} = 19.8$).
Note that none of the four has a counterpart
either in the XMM-Newton X-ray catalogue \citep{ueda2008}
with a detection limit of
$f(0.5$ -- $2$ keV$) = 6 \times 10^{-16}$ erg cm$^{-2}$ s$^{-1}$,
or in the VLA radio catalogue \citep{simpson2006}
with a detection limit of $f(1.4$ GHz$) = 100\mu$Jy.
All four are out of the SCUBA Half-Degree Extragalactic Survey \citep{coppin2006}.

We cannot completely exclude the possibility that the MIPS photometry
of R4 suffers from confusion,
because there are two objects near R4 in the optical wavebands
with a separation of about $1.8$ and $3.8$ arcsec, respectively.
In the following, we assume that possible source confusion does not
significantly affect the MIPS photometry.

We use the template SEDs of starburst galaxies given in
\cite{lagache2003}\footnote{\texttt{http://www.ias.u-psud.fr/irgalaxies/model.php{\#}SED}}
and \cite{vega2008}\footnote{\texttt{http://adlibitum.oat.ts.astro.it/silva/grasil/}\\\texttt{modlib/modlib.html}}
to estimate SFRs from the observed $24 \mu$m flux density. 
As will be shown below, the SFRs thus estimated have considerably
large uncertainties because $24 \mu$m corresponds to near the blue end 
of dust emission for our objects. 
We use one template ($L_{\rm IR} [L_\odot] = 10^{12}$) from \cite{lagache2003},
and templates of 14 starburst-dominated ULIRGs\footnote{
IR 12112$+$0305, UGC 9913, IR 10565$+$2448, IZW 107, IR 10173$+$0828,
Arp 299, UGC 4881, IC 1623, UGC 8387, UGC 2369,
IIIZW 35, IC 5298, Arp 148, UGC 6436.
}
from \cite{vega2008}.
The range of their $\log \left( L_{\rm IR} [L_\odot]  \right)$ is from $11.39$ -- $12.24$.
We redshift them at $z=3.1$ and $3.7$
and find scaling relations between
$m_{24 \mu {\rm m}}$ and $L_{\rm IR}$ for individual templates.
We then derive SFRs from individual $L_{\rm IR}$
using the relation between SFR$_{\rm IR}$ and $L_{\rm IR}$
given in \cite{kennicutt1998}:
\begin{equation}
{\rm SFR}_{\rm IR} \, [M_\odot {\rm yr}^{-1}]
      = 1.7 \times 10^{-10} L_{\rm IR} [L_\odot].
\label{eq:LIR-SFR}
\end{equation}
We obtain SFR$_{\rm IR}$ $\simeq 4.2 \times 10^3$ -- $2.0 \times 10^4$ [$M_\odot$yr$^{-1}$]
(median SFR$_{\rm IR}$ $\simeq 7.4 \times 10^3$ [$M_\odot$yr$^{-1}$] ) for R4 and
upper limits for the remaining three of
$1.1 \times 10^3$ -- $4.3 \times 10^3$ [$M_\odot$yr$^{-1}$]
(median $\simeq 1.5 \times 10^3$ [$M_\odot$yr$^{-1}$]) for $z=3.1$,
and
$2.8 \times 10^3$ -- $1.3 \times 10^4$ [$M_\odot$yr$^{-1}$]
(median $\simeq 4.9 \times 10^3$ [$M_\odot$yr$^{-1}$]) for $z=3.7$. 
The variation of SFR$_{\rm IR}$  among the $15$ templates 
is thus a factor $\simeq 4$.

\begin{figure*}
 \begin{center}
      \includegraphics[scale=0.92]{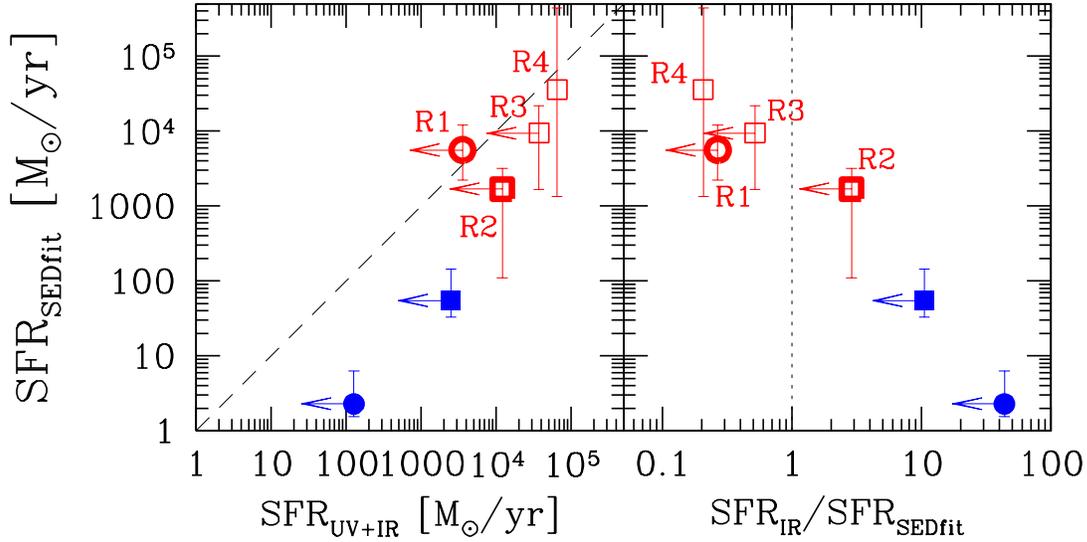}
 \end{center}
 \caption[]{
\textbf{Left}:
Comparison of the SFR derived from the SED fitting with
the SFR derived from the observed UV+IR luminosities.
The dashed line corresponds to SFR$_{\rm SED fit} =$ SFR$_{{\rm UV}+{\rm IR}}$.
The red symbols are the four red LAEs.
The blue filled circle and square are
the $K$-undetected LAEs at $z=3.1$ and $z=3.7$, respectively.
\textbf{Right}:
Fraction of obscured SFR (SFR$_{\rm IR}$/SFR$_{\rm SEDfit}$).
plotted against SFR$_{\rm SEDfit}$.
The dotted line is the equality line.
All symbols are the same as in the left panel.
}
\label{fig:SFR_fit}
\end{figure*}

In the left panel of Figure \ref{fig:SFR_fit},
we compare the SFR derived from the SED fitting (SFR$_{\rm SEDfit}$)
with the sum of the SFRs derived from
the observed UV and IR luminosities (SFR$_{\rm UV+IR}$).
We do not find inconsistency between SFR$_{\rm SEDfit}$ and
SFR$_{\rm UV+IR}$, although the SFR$_{\rm UV+IR}$ of three objects
is an upper limit and thus the constraint on the consistency is
not so strong.
R4 is detected at $24\mu$m and its SFR$_{\rm UV+IR}$ is
comparable to those of SMGs and DRGs.
These results support our inference that
the four red LAEs are massive, dusty starburst galaxies.
They can be detected with SCUBA2/JCMT and Herschel.

In the right panel of Figure \ref{fig:SFR_fit},
the ratio of SFR$_{\rm IR}$ to SFR$_{\rm SEDfit}$ is shown.
The dotted line corresponds to SFR$_{\rm total} =$ SFR$_{\rm IR}$,
i.e., all of the intrinsic UV emission is re-emitted from dust.
Only R4 is detected at $24\mu$m
and has SFR$_{\rm IR}$/SFR$_{\rm total}$ $\simeq 0.2$,
while the other three have upper limits of $0.5$ -- $3$.
This result may suggest that the contribution of the obscured
star formation to the total star formation is not
necessarily very large.

We make median-stacked MIPS $24 \mu$m images
of the LAEs fainter than $K(3\sigma)$ at each redshift
($211$ for $z=3.1$ and $70$ for $z=3.7$).
We find that both are fainter than $m_{24 \mu{\rm m}}(3\sigma)$,
obtaining upper limits of SFR$_{\rm IR} =$
$1.0 \times 10^2$ ($5.8 \times 10^2$) [$M_\odot$yr$^{-1}$]\footnote{
These are median values of the cases with the 15 SB-dominated ULIRGs.
The range of the upper limits of SFR$_{\rm IR}$ are
$73$ -- $2.9 \times 10^2$ [$M_\odot$yr$^{-1}$] for $z=3.1$,
and $3.3 \times 10^2$ -- $1.6 \times 10^3$ [$M_\odot$yr$^{-1}$] for $z=3.7$.
}
for the $z=3.1$ ($3.7$) LAEs.
The results are plotted in blue symbols in Figure \ref{fig:SFR_fit}.
We cannot place strong constraints either on the consistency
between SFR$_{\rm SEDfit}$ and SFR$_{\rm UV+IR}$ (left panel)
or on the fraction of obscured star formation (right panel).

Note that we cannot completely exclude AGN-contaminated objects
from our sample.
In particular, the MIPS detection for R4 may imply that
it has a significant amount of AGN contribution.
In this case, the SFRs calculated above are overestimated.

\subsubsection{Velocity Widths of Lyman $\alpha$ Emission}

As described in Section \ref{subsec:spectroscopy},
we performed spectroscopy of two red LAEs (R1 and R2)
with IMACS on Magellan,
and confirmed them to be real LAEs.
We obtain a Ly$\alpha$ line width of
$v_{\rm FWHM} = 205 \pm 129$ km s$^{-1}$ for R1,
and $v_{\rm FWHM} = 629 \pm 201$ km s$^{-1}$ for R2,
after correction for instrumental broadening
on the assumption that their intrinsic profiles are Gaussian.
\cite{venemans2005} have reported that
the typical FWHM of the Ly$\alpha$ line of LAEs
is $340$ km s$^{-1}$ \citep[see also][]{taniguchi2005,kashikawa2006,tapken2007}.
The velocity width of R1 is similar to this value.
On the other hand, R2 has twice as large a value,
and it might suggest
that this object involves gas outflows \citep[e.g.,][]{taniguchi2000},
that it is a merger of two objects with a large relative velocity,
or that it has a heavily obscured AGN.

\subsubsection{Comparison with Other Red LAEs and Local Starburst
Galaxies}

\begin{figure*}
 \begin{center}
      \includegraphics[scale=0.92]{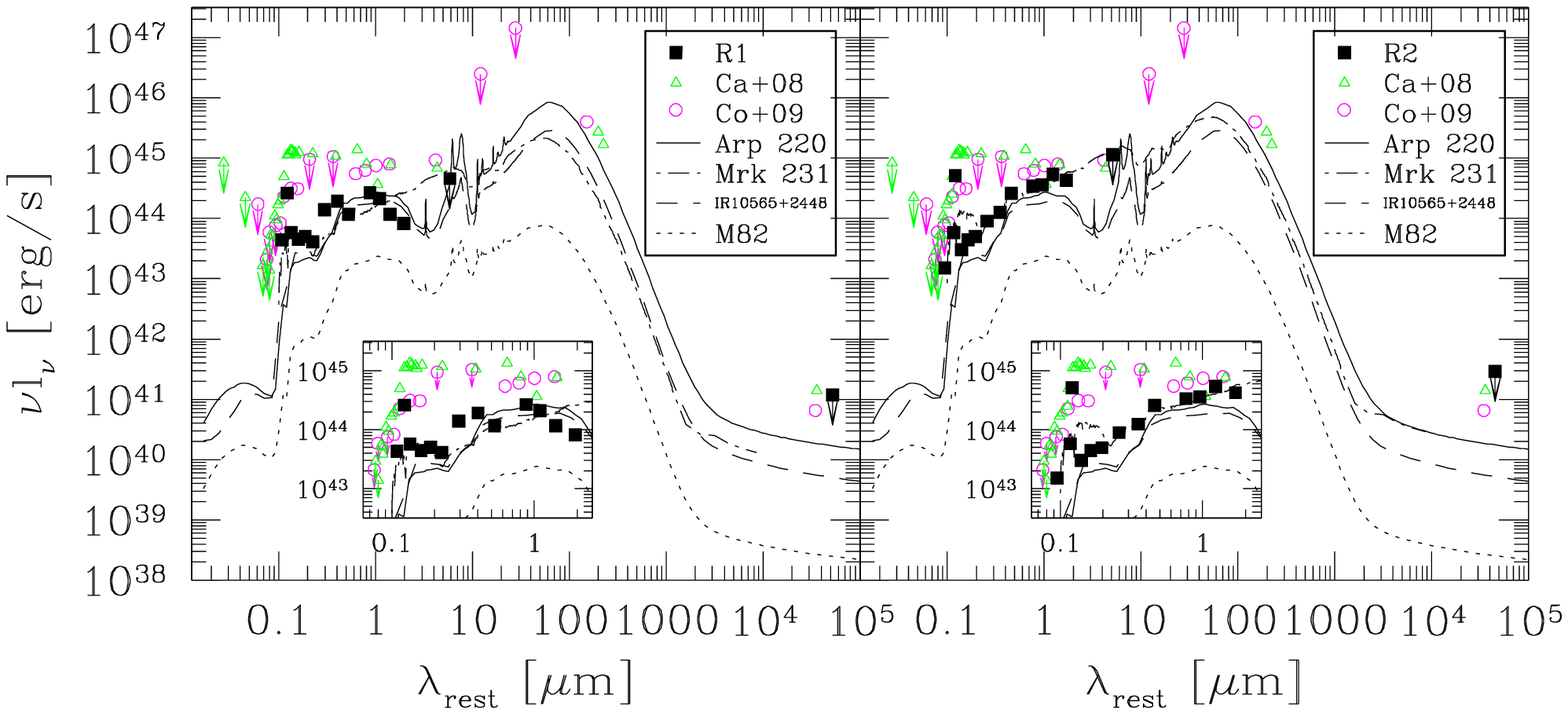}
 \end{center}
 \caption[]{
Multiwavelength SEDs of two spectroscopically
confirmed red LAEs (black squares):
NB503-N-21105 (R1; left) and NB570-N-32295 (R2; right).
Also plotted are the SEDs of
Arp 220 (solid line), M82 (dotted line),
both from \cite{silva1998},
IR10565+2448 (dashed line) from \cite{vega2008},
Mrk 231 (chain line) from \cite{polletta2007}, 
Ly$\alpha$-emitting SMGs
at $z=4.87$ \cite[][magenta circles]{coppin2009}
and at $z=4.547$ \cite[][green triangles]{capak2008}.
The inner panels
are a zoom up over the wavelength range from UV to near-infrared.
}
\label{fig:multiwave_SEDs}
\end{figure*}

We compare in Figure \ref{fig:multiwave_SEDs}
the SEDs of R1 and R2 (both have a spectroscopic redshift)
with those of the three local starburst galaxies
Arp 220, M82 \citep{silva1998}, and
IR10565+2448, which has no AGN contribution to
its IR luminosity \citep{vega2008},
and Mrk 231 \citep{polletta2007},
which is considered to be a local starburst galaxy with a heavily obscured AGN.
Also plotted are two SMGs with strong Ly$\alpha$ emission
recently discovered by \cite{capak2008} and \cite{coppin2009}.

Over the rest-frame UV to NIR wavelengths,
R1 resembles Arp 220 and IR10565+2448 more closely
while R2 resembles Mrk 231, and both are almost one order of
magnitude brighter than M82.
This might indicate that 
R1 is a pure star-forming system, 
while R2 is a star-forming galaxy with an obscured AGN. 
The SEDs of R1 and R2 
including upper limits at the MIPS and VLA wavelengths are
consistent with them being dusty starburst galaxies.
The two SMGs are even brighter than our objects, Arp 220, Mrk 231,
and IR10565+2448,
and have significantly bluer SEDs.

\begin{table*}
\centering
\caption{Comparison of stellar populations of dusty starburst galaxies}\label{tab:DustySB}
\begin{tabular}{ccccccc}
\hline
Object
  & $M_{\rm star}$
  & $E(B-V)_\star$
  & Age
  & SFR
      & Reference
      & Remark
\\
  & $[M_\odot]$
  & [mag]
  & [Myr]
  & [$M_\odot$ yr$^{-1}$]
      &
      &
\\ \hline
NB503-N-21105 (R1)
 & $2.7 \times 10^{10}$
 & $0.70$
 & $4.8$
 & $5.6 \times 10^3$
 & This Study
 &
\\
NB570-N-32295 (R2)
 & $4.0 \times 10^{10}$
 & $0.67$
 & $25$
 & $1.7 \times 10^3$
 & This Study
 &
\\
Arp 220
 & $2.5 \times 10^{10}$
 & $0.6$
 & ($50$)
 & $5.8 \times 10^2$
 & \cite{silva1998}
 & 1
\\
M 82
 & $2 \times 10^{8}$
 & $0.28$
 & ($50$)
 & $5.5$
 & \cite{silva1998}
 & 1
\\
IR10565+2448
 & $2.3 \times 10^{10}$
 & --- 
 & $81$
 & $281.5$
 & \cite{vega2008}
 & 2
\\
$z=4.87$ SMG
 & $\la 5 \times 10^{10}$
 & $0.37$
 & $40$
 & $5\times 10^2 $
 & \cite{coppin2009}
 & 3
\\
$z=4.55$ SMG diffuse
 & $2 \times 10^{10}$
 & $0.37$
 & $7.6$
 & $4.2 \times 10^3 $
 & \cite{capak2008}
 & 4
\\
\hline
\end{tabular}

\medskip
\begin{minipage}{140mm}
\begin{flushleft}

$^1$
They have used a library generated with
GRASIL \citep{silva1998} to make model SEDs.
The stellar mass shown here is the gas mass which was converted
into stars during the most recent burst (see below).
We calculate the colour excess $E(B-V)_\star$
from the optical depth in the $B$ band they have reported
($\tau_B \sim 2.8$ for Arp 220, and
$\tau_B \sim 1.3$ for M82),
using \cite{calzetti2000} extinction curve.
For both objects, they assumed that the SFR increased
in the first $3$ Gyr and then smoothly decreased until
an age of 13 Gyr, followed by a starburst which occurred at
$5 \times 10^7$ years ago.
The SFRs shown here are the ones averaged
over the last $5 \times 10^7$ years.

$^2$
They have used a library generated with
GRASIL \citep{silva1998} to make model SEDs.
The assumed an exponentially decaying starburst.
For age and stellar mass,
we adopt the age of the burst (age$_{\rm b}$)
and the mass of stars created during the burst ($M^b_\star$).
The SFR we adopt is the one averaged over the period of the burst.

$^3$
They have used
the stellar population model of \cite{bc1993},
the \cite{calzetti2000} extinction law,
and either a constant or single-burst star formation history,
assuming $Z=Z_\odot$.
They have derived its stellar mass
from an upper limit of
the rest-frame $K$-band luminosity
and the mass-to-luminosity ratio
obtained by \cite{borys2005}.
They have reported
the dust extinction amount in the $V$ band to be
$A_{\rm V} \sim 1.5$.
We calculate its colour excess $E(B-V)_\star$
using \cite{calzetti2000} law.
They have estimated the SFR
by dividing the stellar mass by
an age of $\sim 100$ Myr.

$^4$
Since this object has a UV bright knot and a diffuse component, 
they have analysed separately the knot 
and the diffuse component. 
We adopt here the values for the diffuse component.
They have used the stellar population model
of \cite{bc2003}, \cite{salpeter1955} IMF,
\cite{calzetti2000} extinction law,
and a single-burst star formation history,
assuming $Z=Z_\odot$.
They have reported
the dust extinction in the $V$ band, and
we calculate the colour excess $E(B-V)_\star$
using \cite{calzetti2000} law.

\end{flushleft}
\end{minipage}
\end{table*}

Table \ref{tab:DustySB} summarises their stellar populations.
R1 and R2 are found to have similar stellar masses
to Arp 220, IR10565+2448, and the two SMGs
(For the $z=4.55$ object we adopt the values for the diffuse component);
M82 is $\sim 100$ times less massive.
The $E(B-V)_\star$ values of R1 and R2 are close to
that of Arp 220,
and the other objects are less obscured.
Ages are not strongly constrained,
but all objects appear to be young.
The SFRs of R1 and R2 are several times higher
than that of Arp 220 and
are comparable to that of the $z=4.55$ SMG.
Thus Table \ref{tab:DustySB} leads us to concluding
that R1 and possibly R2 are high-$z$ counterparts of present-day ULIRGs.

The number density of the red LAEs (R1 -- R4) in our sample
is $\sim 2 \times 10^{-6}$ Mpc$^{-3}$ at $z=3.1$,
and $\sim 7 \times 10^{-6}$ Mpc$^{-3}$ at $z=3.7$, respectively.
These values are about one order of magnitude lower
than that of SMGs at $z \sim 2$ -- $3$
\citep{chapman2005}.
If red LAEs and SMGs are both massive dusty starburst galaxies,
then the difference in the number density would imply that
dusty starburst galaxies emit Ly$\alpha$ photons
in only a short time in their starburst phase.
In that time the distribution of dust in the galaxy
might be clumpy so that Ly$\alpha$ photons can escape
relatively easily.

Note that we cannot rule out the existence of
obscured AGNs in our red LAEs.
Actually, \cite{geach2009} have reported that
all AGN in the Ly$\alpha$ blobs
in the $z=3.09$ proto-cluster in the SSA22 field
appear to be heavily obscured.
If the contribution from AGNs to the SEDs is significant,
our discussion about their stellar populations
will have to be modified.
Even in such cases, however, our finding of R1 -- R4
demonstrates that Ly$\alpha$ surveys can detect
galaxies with very red UV-to-optical SEDs,
which are difficult to be identified
by other photometric selection methods.

\subsection{Contribution of LAEs to the Stellar Mass Density
and the Cosmic Star Formation Rate Density}

We estimate the contribution from LAEs
(both $K$-detected and $K$-undetected)
to the cosmic stellar mass density at $z=3.1$ and $3.7$,
by dividing the total stellar mass of LAEs
by the comoving volume searched by each narrow band.
The total stellar mass at each redshift is defined as the stellar
mass of the stacked LAEs multiplied by their number, plus the sum
of the $K$-detected LAEs' masses.
We obtain
$1.4 \times 10^5$[$M_\odot$Mpc$^{-3}$] for $z=3.1$ LAEs,
and
$5.2 \times 10^5$[$M_\odot$Mpc$^{-3}$] for $z=3.7$ LAEs.
These values can be regarded as lower limits of the contribution
to the cosmic stellar mass density.

\cite{grazian2007} have obtained the stellar mass densities of LBGs
and DRGs at $z \sim 3$ by integrating their stellar
mass functions, to be
$\sim 1 \times 10^7$[$M_\odot$Mpc$^{-3}$] for LBGs,
and
$\sim 7 \times 10^6$[$M_\odot$Mpc$^{-3}$] for DRGs.
The stellar mass densities of our LAEs are 1 -- 3 $\%$ of that of
LBGs$+$DRGs.

Similarly, we estimate the contribution from LAEs
(both $K$-detected and $K$-undetected)
to the cosmic star formation rate density,
where the total star formation rate
is computed as the star formation rate of the stacked objects
multiplied by their number, plus the sum of the $K$-detected
LAEs' star formation rates.
We obtain
$8.2 \times 10^{-3}$[$M_\odot$ yr$^{-1}$Mpc$^{-3}$] for $z=3.1$ LAEs,
and
$1.3 \times 10^{-1}$[$M_\odot$ yr$^{-1}$Mpc$^{-3}$] for $z=3.7$ LAEs.
The cosmic star formation density due to all galaxies
is estimated to be
$\log \dot{\rho}_* \sim 0.18 $ [$M_\odot$ yr$^{-1}$Mpc$^{-3}$] at $z=3.1$,
and
$\log \dot{\rho}_* \sim 0.14 $ [$M_\odot$ yr$^{-1}$Mpc$^{-3}$] at $z=3.7$
\citep{hopkins2006}.
Thus the lower limit to the contribution from LAEs is
about $0.04$ -- $1$ {\%}.

It is interesting to note that for our LAEs both the cosmic star
formation rate and the stellar mass density are higher at $z=3.7$.
This may mean that at higher redshift a higher fraction of galaxies
are actively forming stars, or that the fraction of galaxies with
Lyman $\alpha$ EWs large enough to be selected as LAEs increases
with redshift.


\section{Summary and Conclusions} \label{sec:sum}

In this paper, we have investigated the stellar populations of LAEs
at $z = 3.1$ and $3.7$ found in 0.65 deg$^2$ of the SXDF, based on
deep, rest-frame UV-to-optical photometry from three surveys made
with Suprime-Cam/Subaru, WFCAM/UKIRT, and IRAC/Spitzer.
Among a total of $302$ LAEs ($224$ for $z=3.1$ and $78$ for $z=3.7$),
only $11$ (or $4${\%}) are detected in the $K$ band, i.e., brighter
than $K(3\sigma) = 24.1$ mag (AB).
Among the $11$ $K$-detected LAEs,
eight are spectroscopically confirmed.
In our stellar population analysis, we treat $K$-detected
objects individually, while we stack $K$-undetected objects
at each redshift to derive an average SED.
Since the vast majority ($96${\%}) are undetected in $K$,
the SEDs constructed from stacking should closely
represent the whole LAE population at each redshift.
We treat the LAEs at $z=3.1$ and $3.7$ collectively as objects
at $z \sim 3$, without discussing possible evolution between
the two redshifts.
Our LAE sample, based on deep optical and near-infrared data
over a wide sky area, enables us not only to place strong
constraints on the average properties of LAEs by stacking of
many objects, but also to study rare, massive LAEs visible in $K$.
We fit stellar population synthesis models to the multiband
photometry of $K$-detected and $K$-undetected (stacked) objects,
to derive their stellar masses, star formation rates, ages, and
dust extinctions.
We assume a constant star formation rate and fix the metallicity
to $Z=0.2Z_\odot$.

Our main results are as follows:

\begin{enumerate}
\item
The $K$-detected objects have stellar masses of
$M_{\rm star} \sim 10^9$ -- $10^{10.5} M_\odot$,
while the $K$-undetected objects are 1 -- 2 orders of magnitude
less massive, with $M_{\rm star} \sim 10^8$ -- $10^{8.5} M_\odot$.

\vspace{5pt}
\item
The $K$-detected objects are distributed in a similar region
in the $M_{\rm star}$ versus $M_{\rm V}$ plane to spectroscopically
confirmed Lyman-break galaxies (LBGs) at $z \sim 3$,
while the $K$-undetected objects have lower $M_{\rm star}/L_{\rm V}$
ratios and bluer $M_{1500} - M_{\rm V}$ colours than the $K$-detected ones
(Figure \ref{fig:R_RK_z3_all} and
the left panel of Figure \ref{fig:restOptUV_Mstar}).

\vspace{5pt}
\item
The star formation rates (SFR) of $K$-detected objects
span a wide range of $10$ -- $10^4 M_\odot$ yr$^{-1}$,
while those of $K$-undetected objects fall
between $1$ -- $100 M_\odot$ yr$^{-1}$.

\vspace{5pt}
\item 
There could be a bimodality in the age distribution
among the $K$-detected objects (Figure \ref{fig:histograms}).
The ages of the $K$-undetected LAEs are around the median value
of the ages of the $K$-detected LAEs.

\vspace{5pt}
\item
Four of the $K$-detected objects have extremely large $E(B-V)_\star$,
and thus their SFRs derived from rest-frame 1500 \AA\ continua and
from Lyman $\alpha$ luminosities are both underestimated by
more than two orders of magnitude
(Figure \ref{fig:restUVLLyA_SFR}).
On the other hand, the dust extinction of $K$-undetected objects
is modest.

\vspace{5pt}
\item
The Lyman $\alpha$ equivalent width (EW) weakly anti-correlates
with $M_{\rm star}$, except for the four dusty objects having
large EWs (Figure \ref{fig:EW_Pops}).
No significant correlation is seen between EW(Ly$\alpha$) and $E(B-V)_\star$
(Figure \ref{fig:EW_Pops}).

\vspace{5pt}
\item
The Lyman $\alpha$ escape fraction decreases with $E(B-V)_{\rm gas}$.
At a fixed $E(B-V)_{\rm gas}$, the escape fraction of our LAEs is higher
than those of LBGs and
local star-forming galaxies (Figure \ref{fig:EBV_fesc}).

\vspace{5pt}
\item
The stellar masses and the specific star formation rates (sSFR)
of LAEs are compared with those of LBGs,
DRGs, SMGs, and galaxies
with photometric redshifts of $z_{\rm phot} \sim 3$.
It is found that the LAE population as a whole is the least
massive among these galaxy populations, but with relatively
high sSFRs (Figure \ref{fig:Mstar_sSFR}).

\vspace{5pt}
\item
Four of our $K$-detected LAEs with the reddest SEDs,
two of which are spectroscopically confirmed,
are heavily obscured with $E(B-V)_\star \sim 0.65$,
and their continua resemble that of Arp 220 and the local ULIRG
(Figure \ref{fig:multiwave_SEDs}).
They have very high sSFRs in spite of their large
stellar masses (Figure \ref{fig:Mstar_sSFR}),
and one of them are detected in the MIPS $24\mu$m image,
which supports the SED-fitting result of SFR.
These facts suggest that the red LAEs are massive, dusty starburst galaxies.
Our study demonstrates that wide-field Lyman $\alpha$ surveys
can detect such dusty starburst galaxies.

\end{enumerate}

\section*{Acknowledgements}

We would like to thank the anonymous referee for very constructive comments and suggestions.
We are also very grateful
to Gustavo Bruzual and St\'ephane Charlot for sending us their new GALAXEV and stellar libraries;
to Peter Capak for telling us the information on the SEDs of their SMGs;
to Kristen Coppin for providing the SED of the Ly$\alpha$-emitting SMG;
to Georg Feulner for telling us the webpage containing their photometric redshift catalogue;
to Steven Finkelstein for sending us the data on the SFRs of their LAEs derived from the best-fit SEDs.
RM would like to acknowledge the funding of the Royal Society.

\vspace{0.5em}

\noindent
\textit{Facilities}:
Subaru (Suprime-Cam),
Magellan: Baade (IMACS),
UKIRT (WFCAM),
Spitzer (IRAC, MIPS)



\label{lastpage}

\end{document}